\documentclass[fleqn,usenatbib]{mnras}

\usepackage{newtxtext,newtxmath}

\usepackage[T1]{fontenc}

\DeclareRobustCommand{\VAN}[3]{#2}
\let\VANthebibliography\thebibliography
\def\thebibliography{\DeclareRobustCommand{\VAN}[3]{##3}\VANthebibliography}

\usepackage{graphicx}	
\usepackage{amsmath}	
\usepackage{amssymb}	
\usepackage{threeparttable}

\newcommand{\unsim}{\mathord{\sim}}

\title[Kuiper Belt Classification]{Machine Learning Classification of Kuiper Belt Populations}

\author[Smullen \& Volk]{
Rachel A. Smullen,$^{1}$\thanks{E-mail: rsmullen@email.arizona.edu}
Kathryn Volk$^{2}$
\\
$^{1}$Department of Astronomy, University of Arizona, 933 N Cherry Ave., Tucson 85719 USA\\
$^{2}$Lunar and Planetary Laboratory, University of Arizona, 1629 E. University Blvd., Tucson 85719 USA
}

\date{Accepted XXX. Received YYY; in original form ZZZ}

\pubyear{2020}

\begin{document}
\label{firstpage}
\pagerange{\pageref{firstpage}--\pageref{lastpage}}
\maketitle

\begin{abstract}

In the outer solar system, the Kuiper Belt contains dynamical sub-populations sculpted by a combination of planet formation and migration and gravitational perturbations from the present-day giant planet configuration. The subdivision of observed Kuiper Belt objects (KBOs) into different dynamical classes is based on their current orbital evolution in numerical integrations of their orbits. Here we demonstrate that machine learning algorithms are a promising tool for reducing both the computational time and human effort required for this classification.  Using a Gradient Boosting Classifier, a type of machine learning regression tree classifier trained on features derived from short numerical simulations, we sort observed KBOs into four broad, dynamically distinct populations--classical, resonant, detached, and scattering--with a >97 per cent accuracy for the testing set of 542 securely classified KBOs. Over 80 per cent of these objects have a $>3\sigma$ probability of class membership, indicating that the machine learning method is classifying based on the fundamental dynamical features of each population. We also demonstrate how, by using computational savings over traditional methods, we can quickly derive a distribution of class membership by examining an ensemble of object clones drawn from the observational errors.  We find two major reasons for misclassification: inherent ambiguity in the orbit of the object--for instance, an object that is on the edge of resonance--and a lack of representative examples in the training set. This work provides a promising avenue to explore for fast and accurate classification of the thousands of new KBOs expected to be found by surveys in the coming decade. 
\end{abstract}

\begin{keywords}
Kuiper belt: general -- planets and satellites: dynamical evolution and stability -- methods: data analysis -- methods: statistical
\end{keywords}



\section{Introduction}

\subsection{The Kuiper Belt}

The Kuiper Belt consists of many sub-populations of small bodies in the outer solar system. 
The orbital distribution of Kuiper Belt objects (KBOs) records the complex early dynamical history of the solar system's giant planets as well as a variety of current dynamical processes \citep[see, e.g.,][]{Morbidelli:2008,Dones:2015}. In order to use observations of KBOs to constrain processes in the current and early solar system, they must be classified into different dynamical groups. {The classification of these populations is a critical first step towards identifying distinct compositional classes in the Kuiper Belt that might be indicative of formation processes \citep[e.g.,][]{Pike:2017} and constraining models of the early dynamical evolution of the solar system by making direct comparisons between models and observations \citep[e.g.,][]{Nesvorny:2015,Nesvorny:2019b,Nesvorny:2019,Chen2019}.}

Different classification schemes exist for the Kuiper Belt, but here we will use \citet{Gladman:2008}'s scheme. 
Briefly, this scheme divides KBOs into classical belt objects, scattering objects, resonant objects, and detached objects, the detailed definitions of which are described in more detail in Section~\ref{ss:classification}. 
Classical belt objects are a mixture of dynamically cold and dynamically excited KBOs mostly in the semi-major axis range $a\sim36-50$~au; identifying these objects in an observational sample is important because the dynamically cold sample of classical objects are generally thought to have formed in situ and to represent a remnant of the original planetesimal disc (see, for example, the recent results about the primordial origins of the classical KBO Arrokoth \citealt{McKinnon:2020}).
Resonant objects are those in mean motion resonances with Neptune.
Resonant objects are of particular importance for constraining the dynamical history of the outer solar system; while some KBOs are merely temporarily resonant \citep[see, e.g.,][]{Lykawka:2007,Yu:2018}, the large number of KBOs in resonance with Neptune is best understood to be a result of resonant capture during the epoch of planetary migration in the early solar system \citep[see, e.g.][]{Morbidelli:2008,Malhotra:2019}. 
Detached objects are those KBOs that (on the 10\,Myr dynamical time-scale of the \cite{Gladman:2008} scheme) do not appear to experience significant orbital evolution, i.e., they are dynamically `detached' from the giant planets. 
These objects are important to identify because their current orbits are difficult to obtain through interactions with the currently known planetary system. 
Truly detached objects must have been dynamically perturbed on to their current orbits {either} by large bodies or {by} processes in the early solar system that are not longer extant (e.g., rogue planets \citealt{Gladman:2006}, close stellar flybys \citealt{Kaib:2011}, interactions within the sun's birth cluster \citealt{Brasser:2012}, resonant drop-out while Neptune migrates \citealt{Lawler:2019}).
Scattering objects are those that are currently being strongly dynamically perturbed by direct gravitational interactions with Neptune, thus reflecting current solar system conditions. 

\subsection{Machine Learning Classification in Astronomy}

Astronomy has long been an ideal field for applications of machine learning--statistical methods that learn to recognize data based on patterns and inference--due to the large data volume and wide range of problem complexity. Beginning with simple clustering and neural network applications for galaxy classification \citep[e.g,][]{Adorf1988,Storrie-Lombardi1992}, astronomers have been adopting more varied and sophisticated machine learning methodologies to utilize the full spectrum of information contained in both observations and simulations. Indeed, the importance of machine learning integration in future astronomy programs 
{has been discussed in many works, such as \cite{Nord2019}. }

Several recent investigations in astronomical machine learning use time-dependent features from the outputs of numerical simulations for classification. For instance, \cite{Tamayo2016} use a standard machine learning classifier and features derived from semi-major axis and eccentricity in short {N-body} simulations to predict the orbital stability of three planet systems. \cite{McLeod2017} use cosmological simulations and a neural network to predict the mass of the Local Group. \cite{Lam2018} train a deep neural network to predict the stability of circumbinary planetary systems using only four dynamical features {derived from N-body simulations}. \cite{Choudhary2019} shows how machine learning methods (in this case, a Hamiltonian neural network) can even learn to predict orbital dynamics (such as long-term chaotic or non-chaotic orbits) without the need for numerical simulations. For dynamical{ly-evolving} systems, a single snapshot in time is insufficient for classification and thus, it is critical to incorporate more complex, time-dependent data for a complete prediction.

\smallskip
A methodology to quickly and accurately classify new objects in dynamical populations is especially critical for solar system purposes. The Vera Rubin Observatory's Legacy Survey of Space and Time (LSST) is expected to find millions of solar system objects, including a few tens of thousands of new KBOs \citep[see, e.g.,][]{Schwamb2018, Ivezic:2019}. Traditional classification procedures for the Kuiper Belt, particularly, require {some level of human intervention/verification for nearly every object}, which quickly becomes unsustainable with ballooning data size. {A more efficient tool for dynamical classification of KBOs detected by LSST will be needed \citep{Schwamb:2019}.}

Thus, in this paper, we demonstrate the efficacy of a machine learning classification algorithm on separating observed KBOs into their component dynamical populations. {This method allows for fast and accurate classification while substantially reducing the need for human intervention.} We describe the data and machine learning algorithm in Section~\ref{s:data}. We then show the results of our classifier on the testing data, including the robustness of our results, in Section~\ref{s:results}. Finally, in Section~\ref{s:discussion}, we discuss the implications of our investigation and explore improvements to the methodology shown herein.

\section{Data and Methods}\label{s:data}

{The goal of a classification scheme is to take the observed orbits of KBOs, run numerical integrations of their evolution, and then classify the objects into dynamical categories based on these integrations. In Section~\ref{ss:classification}, we describe the set of KBOs used as the training and testing data for the machine learning classifier; these KBOs have been previously classified according to the \cite{Gladman:2008} scheme. Section~\ref{ss:ml_method} then describes the training and refinement of the machine learning algorithm for classifying the KBOs.}

\subsection{Kuiper Belt Observations and Classification}\label{ss:classification}

Here we describe the data set of observed KBOs used in this paper, which is the set of KBOs that was examined in \citet{Volk:2017} (all multi-opposition objects beyond Neptune available in the Minor Planet Center\footnote{\url{www.minorplanetcenter.net}} database as of late 2016) combined with the set of classified objects from the Outer Solar System Origins Survey reported by \citet{Bannister:2018}.
The classifications of these $\unsim2300$ KBOs were produced following the procedures detailed in \citet{Gladman:2008}. 
Briefly, an orbit is fit to the observations of each object using the \citet{Bernstein:2000} orbit-fitting code.
Then, the uncertainty in that orbit fit is estimated by finding the largest deviations in semi-major axis on either side of the best-fitting orbit that does not produce observational residuals that are more than 1.5 times worse than the residuals from the best-fitting orbit. 
This produces three versions, or `clones' of the observed KBO that are integrated forward in time under the gravitational influence of the Sun and the four giant planets (using SWIFT; \citealt{Levison:1994}) for 10\,Myr. 

For objects with semi-major axes beyond Neptune, the dynamical evolution of each clone {(as determined by the time evolution of barycentric orbital elements)} is then classified into the following categories. 

\begin{enumerate}

\item Resonant objects show libration of a mean motion resonance argument for {more than 50 per cent} of the 10\,Myr span. Mean motion resonances occur when a KBO's orbital period is commensurate with Neptune's orbital period, which results in the libration (rather than circulation) of some combination of the KBO's mean longitude $\lambda_\textrm{KBO}$, Neptune's mean longitude $\lambda_N$, and the KBO's longitude of perihelion $\varpi_\textrm{KBO}$ and/or longitude of ascending node $\Omega_\textrm{KBO}$ (see \citealt{Murray1999} for a complete description of mean motion resonances); an example is Neptune's 3:2 resonance {in which Pluto-Charon resides}, for which the angle $\phi = 3 \lambda_\textrm{KBO} - 2\lambda_N - \varpi_\textrm{KBO}$ librates around $\phi = 180^\circ$.

\item Scattering objects are objects whose semi-major axes $a$ change by more than 1.5~au over the course of the 10\,Myr integration. This is a result of gravitational interactions with Neptune that change the energy (and therefore semi-major axis) of the objects' orbit. In practice, most scattering objects have perihelion distances $q\lesssim37-38$~au, the rough boundary where strong direct interactions with Neptune are possible at perihelion \citep[e.g.,][]{Gladman:2002}. 

\item Detached objects are objects with large eccentricities ($e>0.24$) but that do not experience significant changes in semi-major axis {($\Delta a < 1.5$~au)} over 10\,Myr. This indicates a lack of strong interactions with Neptune; detached objects typically have semi-major axes $a\gtrsim 50$~au and large perihelion distances. 

\item Finally,
classical objects are KBOs that do not fall into any of the above categories. Objects in the `main' classical belt have semi-major axes between the 3:2 and 2:1 resonances with Neptune {($39\textrm{\,au} < a < 48\textrm{\,au}$)}; 'inner' classical objects fall interior to the 3:2, and `outer' classical objects are exterior to the 2:1. (See \citealt{Gladman:2008} for full discussion of the motivations for these classification boundaries).
\end{enumerate}
{These classifications are determined by a combination of simple time-series analysis and visual inspection of the orbital evolution.}

Table~\ref{T:cat} breaks down how our $\unsim2300$ KBO data set is divided between these four categories. The table is divided into secure and insecure classifications; secure classifications are those objects for which all three clones behave similarly during the 10\,Myr integrations while insecure objects are those with differing classifications between clones of the same object. 
Insecure classifications are often the result of large uncertainties in the orbit of a KBO, although sometimes it reflects an object with a very well-determined orbit being near the boundary of dynamical classes (often being very near the edge of one of Neptune's mean motion resonances). {We use only the $\sim1800$ securely classified objects in our training and initial assessment of machine learning algorithms (Section~\ref{ss:ml_method} and Section~\ref{s:results}); we discuss how the classifier performs on the insecure objects in Section~\ref{sss:insecure}.}

\begin{table}
\caption{Overview of Observed KBO Catalog} 
\label{T:cat}
\centering
\begin{tabular}{lrr}
\hline
Population  & Secure & Insecure  \\
\hline
\textbf{All} & \textbf{1805} & \textbf{500} \\
\hline
\textbf{Resonant} & \textbf{642} & \textbf{184} \\
\hspace*{0.5cm} 3:2 & 333 & 14 \\
\hspace*{0.5cm} 2:1 & 71 & 6 \\
\hspace*{0.5cm} 7:4 & 51 & 31 \\
\hspace*{0.5cm} 5:2 & 48 & 7 \\
\hspace*{0.5cm} 5:3 & 45 & 11 \\
\hspace*{0.5cm} 4:3 & 26 & 3 \\
\hspace*{0.5cm} 3:1 & 11 & 9 \\
\textbf{Classical} & \textbf{998} & \textbf{151} \\
\hspace*{0.5cm} Main & 941 & 139 \\
\hspace*{0.5cm} Inner & 43 & 6 \\
\hspace*{0.5cm} Outer & 14 & 5 \\
\textbf{Detached} & \textbf{74} & \textbf{90} \\
\textbf{Scattering} & \textbf{91} & \textbf{75} \\
\hline
\end{tabular}

\end{table}

\subsection{Classifier Selection and Training} \label{ss:ml_method}

A supervised machine learning classifier, such as we use here, must have labelled (pre-classified) data to train and test upon so that we can calculate a method accuracy. \emph{Thus, we make the critical assumption that the \cite{Gladman:2008} 10\,Myr classification of KBOs described above represent the `true' class of the object {for our chosen classification scheme.} }{There are other possible classification schemes for observed objects \citep[e.g.,][]{Elliot:2005}; the approach we take here should, in principle, be generalizable to any classification methodology.}

\subsubsection{Data Features}\label{sss:features}

A machine learning classifier is trained on \emph{features}, or properties of an object. For this classification problem, we compute various statistics of the numerical integrations {generated for the classification procedure described in Section~\ref{ss:classification}; these statistics are then used by the classifier} 
to identify the true KBO population that an object belongs to. The simulations output barycentric semi-major axis $a$, eccentricity $e$, inclination $i$, argument of pericentre $\omega$, longitude of the ascending node $\Omega$, and mean anomaly $M$ at fixed time intervals (which, for the fiducial simulations, occur every thousand years). 
For the purposes of this paper, we discard $M$ 
{because this angle (for a fixed orbit) simply varies linearly in time and its evolution depends only on semi-major axis.} 
We take subsamples of the data from time $t=0$ to a range of final times and record the initial, final, minimum, maximum, mean, standard deviation, and maximum deviation of the {remaining five osculating orbital elements}. We also take step-wise time derivatives {at each simulation output} and calculate the minimum, maximum, mean, and maximum deviation of the time derivatives. {We do not normalize any of the features.} {In total,} we compute 11 features for each of the five orbital elements, leading to a total of 55 features per object used in our classification. {To be explicit, every clone of an observed KBO (e.g., best fit orbit, minimum or maximum orbit from observed errors) is treated as an independent object.} While many of the features may be highly correlated, the nature of the classifier used herein (described in Section~\ref{ss:classifier}) reduces the risk of overfitting due to these correlations.

The simulations used for the current classification methodology run to 10\,Myr.
{However, many of the dynamical signatures indicative of class membership can be seen on shorter time-scales (for example, single libration cycles for many of Neptune's mean motion resonances are $\unsim10^4$ years).}
Thus, we explore classification accuracy along the time axis. We output features with final times of $5\times10^3, 10^4, 5\times10^4, 10^5, 5\times10^5, 10^6, 5\times10^6,$ and $10^7$ years. 
{If shorter numerical integrations are sufficient to {securely classify the majority of} objects, then it would be possible to leverage the same computational power per object to more thoroughly explore the uncertainty range in the orbit fits.} 
{Because of the evolving nature of KBO orbits, the 10\,Myr classification may not actually describe the orbital behavior passed to the classifier. We discuss this further in Section~\ref{ss:reasons}.}

\subsubsection{Choosing and Refining a Classifier}\label{ss:classifier}

{Machine learning classification algorithms have been developed and optimized for a multitude of purposes. Each has a type of data that it will classify most accurately. Thus, it is important to test a variety of classifiers to find the best tool for these data.}
{We therefore test 15 of the multi-class classifiers available in \texttt{scikit-learn} \citep{Pedregosa2011a} }with default parameters.\footnote{For transparency and reproducibility, we use a seed of 30 for all instances of a random number generator.} First, we test the support vector machine classifiers: Support Vector Classifier (SVC) and Linear Support Vector Classifier (LSVC). Support vector machines try to create classes that are as well separated as possible in the multi-dimensional feature space;
{these classifiers work best when classes are more discrete rather than more continuous.}
The next ensemble of classifiers are the tree classifiers, which include Gradient Boosting Classifier (GBC), Random Forest Classifier (RFC), AdaBoost Classifier (ABC), Extra Trees Classifier (ETC), and Decision Tree Classifier (DTC). Tree classifiers use many layers of binary classifications based on features to sort data into classes {and are commonly the choice for the type of multi-class classification problem explored herein}. 
{We also test} linear model classifiers, which classify based on a linear combination of features, include Logistic Regression (LR), Passive Aggressive Classifier (PAC), Ridge Classifier (RC), and SGD Classifier (SGDC); Quadratic Discriminant Analysis (QDA) is a similar classifier that uses a quadratic instead of linear decision function {(the function that returns a binary true or false classification for an object)}. The last few classifiers {tested} are Gaussian NB (GNB; based on Bayes' theorem), K-Neighbors Classifier (KNC; computing classes based on clustering {in high-dimensional feature space}), Gaussian Process Classifier (GPC; based on Laplace approximation), and Multi-layer Perceptron classifier (MLPC; a simple neural network classifier).

\begin{figure}
\centering
\includegraphics[width=\columnwidth]{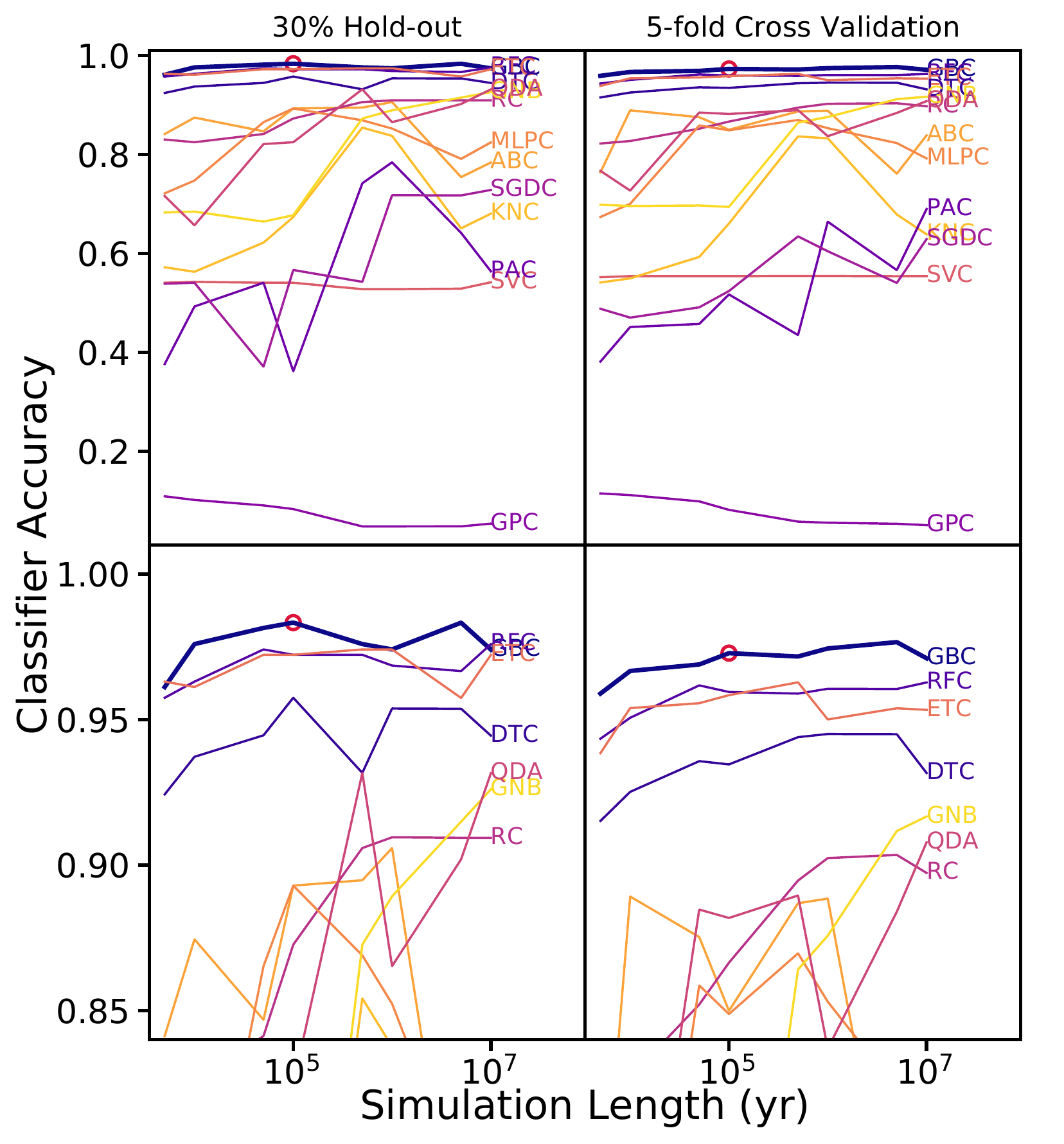} 
\caption{ The accuracy of different {machine learning} classifiers as a function of simulation length. The bottom panels zoom in on the upper 15 per cent of the upper panels. The left panels show accuracy when a random 30 per cent of the data are held out for testing, while the right panels show the accuracy for 5-fold cross validation, in which a random 20 per cent of the data is held out for each of 5 iterations and the final accuracy is the averaged accuracy of all iterations. Every line is labelled with the acronym provided in the text. The best combination of accuracy and simulation length is achieved with the Gradient Boosting Classifier (GBC; thick blue line) at $10^5$ yr, which is marked with a small red circle in each panel.
\label{f:test}}
\end{figure}

We show the accuracy of the different classifiers in Figure~\ref{f:test}.
{The left panels show the accuracy of classifiers trained on a 30 per cent hold-out, meaning that a random 30 per cent of the data are reserved for testing the accuracy of the method trained on the other 70 per cent of the data. While objects in the 30 per cent hold-out testing set are randomly selected from the full ensemble of data, we use the same testing set throughout this work.}
The right panels of Figure~\ref{f:test} show the classifier accuracy in a 5-fold cross validation, meaning that the data are split into 5 subsections. New initializations of the classifier are trained on four subsections and tested on the fifth for all combinations of the subsections. The reported accuracy is then the average accuracy of the 5 different iterations of the classifier. 
Two of the classification methods, LSVC and LR, did not converge (the required accuracy was not reached in a reasonable number of iterations) and are therefore not shown in the figure.
The poorest performing algorithm shown is GPC;
GPC is susceptible to overfitting (fitting specific features in the training set instead of the broad populations) in multi-class classification when different classes occupy similar parameter spaces and therefore shows poor accuracy with the testing set. Most of the tree classifiers performed exceedingly well across all time slices, achieving $>90$ per cent accuracy without further refinement.

{Based on these results, we} choose the Gradient Boosting Classifier (GBC) as the algorithm used for classification throughout the rest of this paper. We also choose {an integration {length} of} 100\,kyr as our fiducial choice: this time subset reaches high accuracy in the classifier ($>97$ per cent without additional refinement), represents more than 100 orbital periods for most distant objects, and {requires a very computationally inexpensive simulation}. 

The next step is to maximize the accuracy of the Gradient Boosting Classifier by optimizing the hyperparameters of the classifier, {which are} the parameters that control how the classifier learns.
A basic description of these parameters and the impact they have on GBC is given in the scikit-learn user guide\footnote{\href{https://scikit-learn.org/stable/modules/ensemble.html\#gradient-tree-boosting}{scikit-learn.org/stable/modules/ensemble.html}.} 
For the purposes of reproducibility, we show the five hyperparameters {we tested} and the range searched in Table~\ref{T:hp};
using all combinations of these variables, we search for the most accurate and efficient combination to use for the rest of this work. {The hyperparameters are validated using a 5-fold cross validation technique.}
{This is a similar approach for hyperparameter optimization as taken in, for example, \citet{Tamayo2016}.}
The top five most accurate combinations of hyperparameters are then given in Table~\ref{T:bestfit}. 
The chosen classifier hyperparameters are given in the bottom line of the table; they create the classifier that has the best combination of speed and accuracy, achieving an average $\unsim98$ per cent accuracy on our data with a 5-fold cross validation (with a maximum accuracy of 99.2 per cent on any individual fold), and a $98.5$ per cent accuracy with the {same 30 per cent hold-out split we use for Figure~\ref{f:test}}. The choice of a small learning rate {(which dampens fluctuations in the algorithm training process)}
leads to a classifier that should be fairly robust against overfitting.

\begin{table}
\caption{Hyperparameter search ranges to optimize method accuracy } 
\label{T:hp}
\begin{threeparttable}
\centering
\begin{tabular}{lll}
\hline
Parameter & Default & Search Range  \\
\hline
Loss\tnote{a} 	        & Deviance & \{Deviance, Exponential\} \\
Learning Rate\tnote{b}	    & 0.1 & \{0.1--1\} with step 0.05 \\
$N_\textrm{estimators}$\tnote{c}& 100 & \{10--500\} with step 10 \\
Maximum Depth\tnote{d} 	      & 3 & \{1--6\} with step size 1 \\
Maximum Features\tnote{e}      & None & \{None, Auto, Sqrt, Log2\} \\
\hline
\end{tabular}

\begin{tablenotes}
\item[a] The function used to quantify the accuracy of the method.
\item[b] The scale of the step length in the gradient descent, which controls overfitting.
\item[c] The number of regression trees in the classifier.
\item[d] The height of the regression tree.
\item[e] The size of the subset of features considered when splitting a node.
\end{tablenotes}

\end{threeparttable}
\end{table}

\begin{table}
\caption{Highest accuracy hyperparameter combinations } 
\label{T:bestfit}
\begin{threeparttable}
\centering
\setlength\tabcolsep{4pt} 
\begin{tabular}{lllllll}
\hline
Rank & Loss & Learning & $N_\textrm{estimators}$ & Max & Max & Accuracy \\
 & Function & Rate & & Depth & Features &  \\
\hline
 1 & Deviance & 0.2 & 140--500\tnote{a} & 3 & Log2 & $97.9\pm0.9$ \\
 2 & Deviance & 0.1 & 130 & 3 & Log2 & $97.9\pm1.1$ \\
 3 & Deviance & 0.15 & 360 & 1 & Log2 & $97.9\pm1.3$ \\
 4 & Deviance & 0.3 & 320 & 1 & Sqrt & $97.9\pm1.4$ \\
 5 & Deviance & 0.15 & 500 & 1 & Log2 & $97.8\pm1.3$ \\
\hline
Best \tnote{b} & Deviance & 0.1 & 130 & 3 & Log2 & $97.9\pm1.1$\\
\hline
\end{tabular}

\begin{tablenotes}
\item[a] All classifiers in this range of estimators gave identical results
\item[b] The best classifier was chosen based on a combination of speed and accuracy.
\end{tablenotes}

\end{threeparttable}
\end{table}

\section{Results}\label{s:results}
In this section, we describe the characteristics of our machine learning classifier tested on the fiducial data set {(100\,kyr simulations of the $\unsim1800$ securely classified KBOs)} with the features described above. We explore the data features that lead to a good classification, study the probabilities of class membership to determine how well our algorithm might perform on unknown data, and investigate the performance of our algorithm on error-space clones of KBOs used in the classifier. {We then explore the performance of the algorithm on other types of objects, such as the insecurely classified observed KBOs, in Section~\ref{s:discussion}. }

\subsection{Object Classification and Feature Importance}\label{ss:classification_features}

\begin{figure}
\centering
\includegraphics[width=\columnwidth]{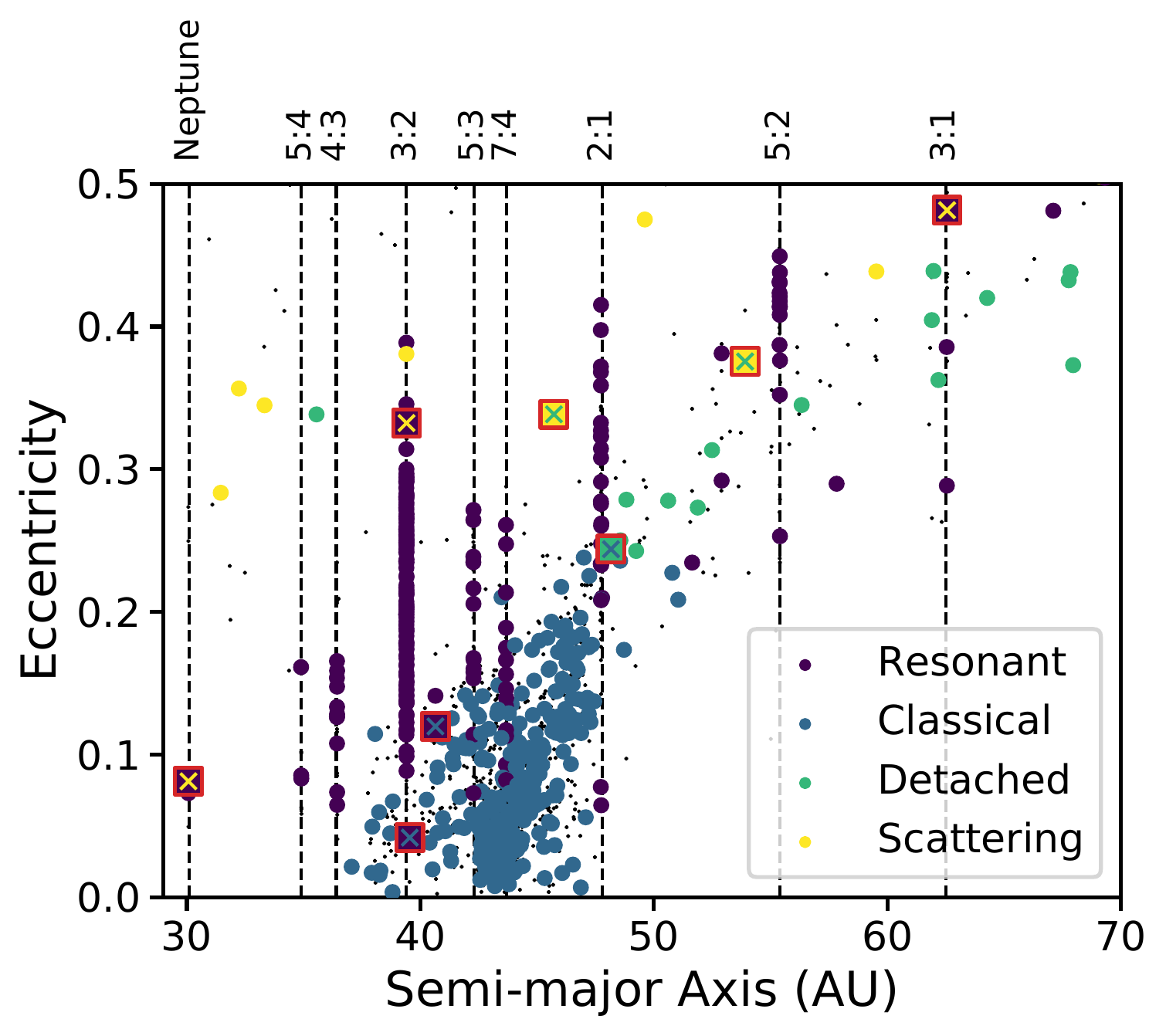} 
\caption{ The classes of objects identified, plotted in the standard orbital plot of eccentricity vs. semi-major axis. The training data are identified as small black points in the background, while the testing data are shown by the larger coloured points. Misclassified objects are identified by red squares, where the background colour in the square shows the true class and the `x' shows the computed class. We achieve a 98 per cent accuracy in our classifier using the 542 objects in our testing set. 
\label{f:4class}}
\end{figure}

\begin{figure}
\centering
\includegraphics[width=\columnwidth]{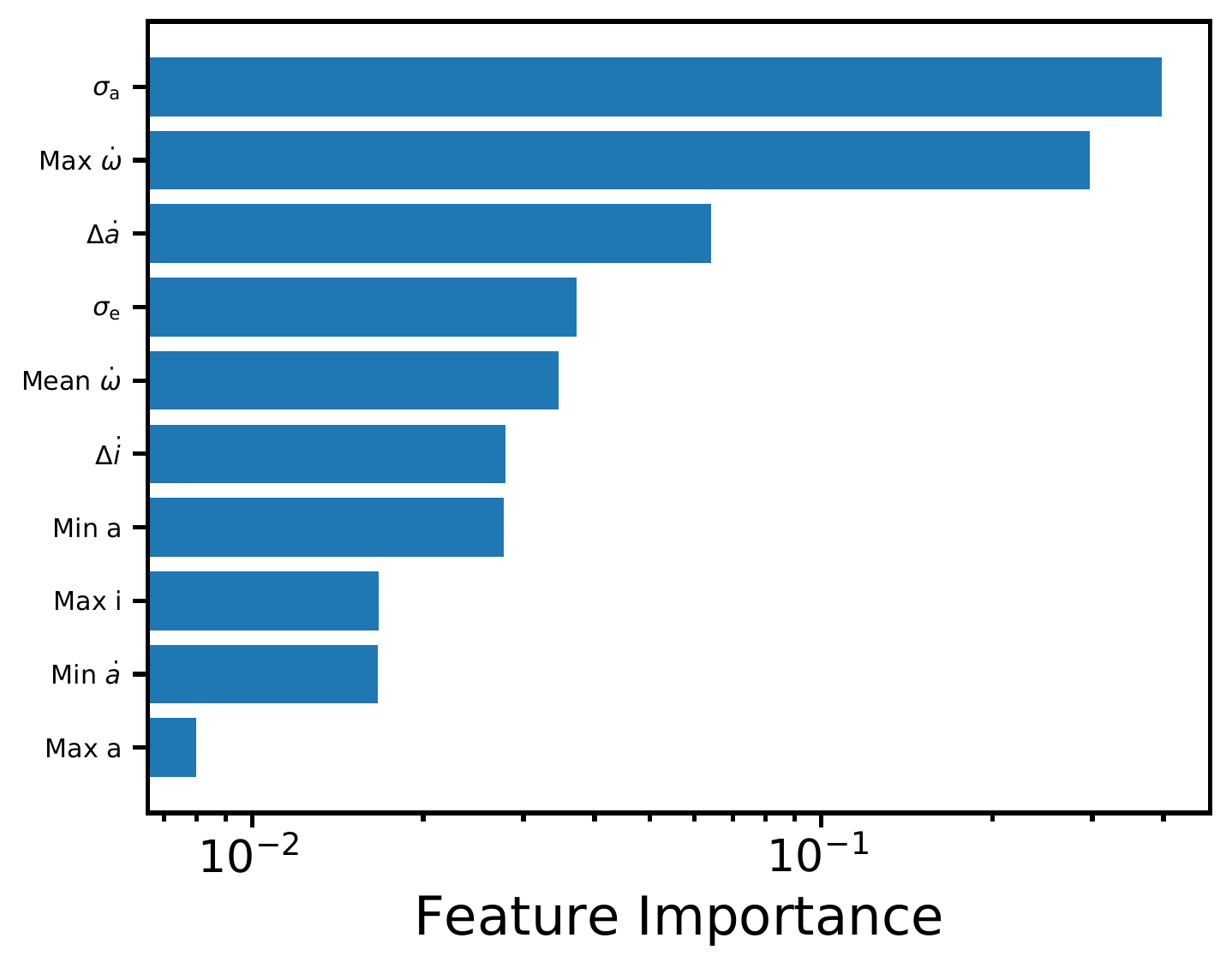} 
\caption{ {Relative importance of the top ten features (from top to bottom: most to least important) used by the classifier to sort objects into their populations}. The sum of all feature importances adds to one. As can be seen, the most important orbital features for classification are the standard deviation of semi-major axis (which will be indicative of scattering or libration) and the maximum rate of change of the argument of pericentre. 
\label{f:features}}
\end{figure}

\begin{figure}
\centering
\includegraphics[width=\columnwidth]{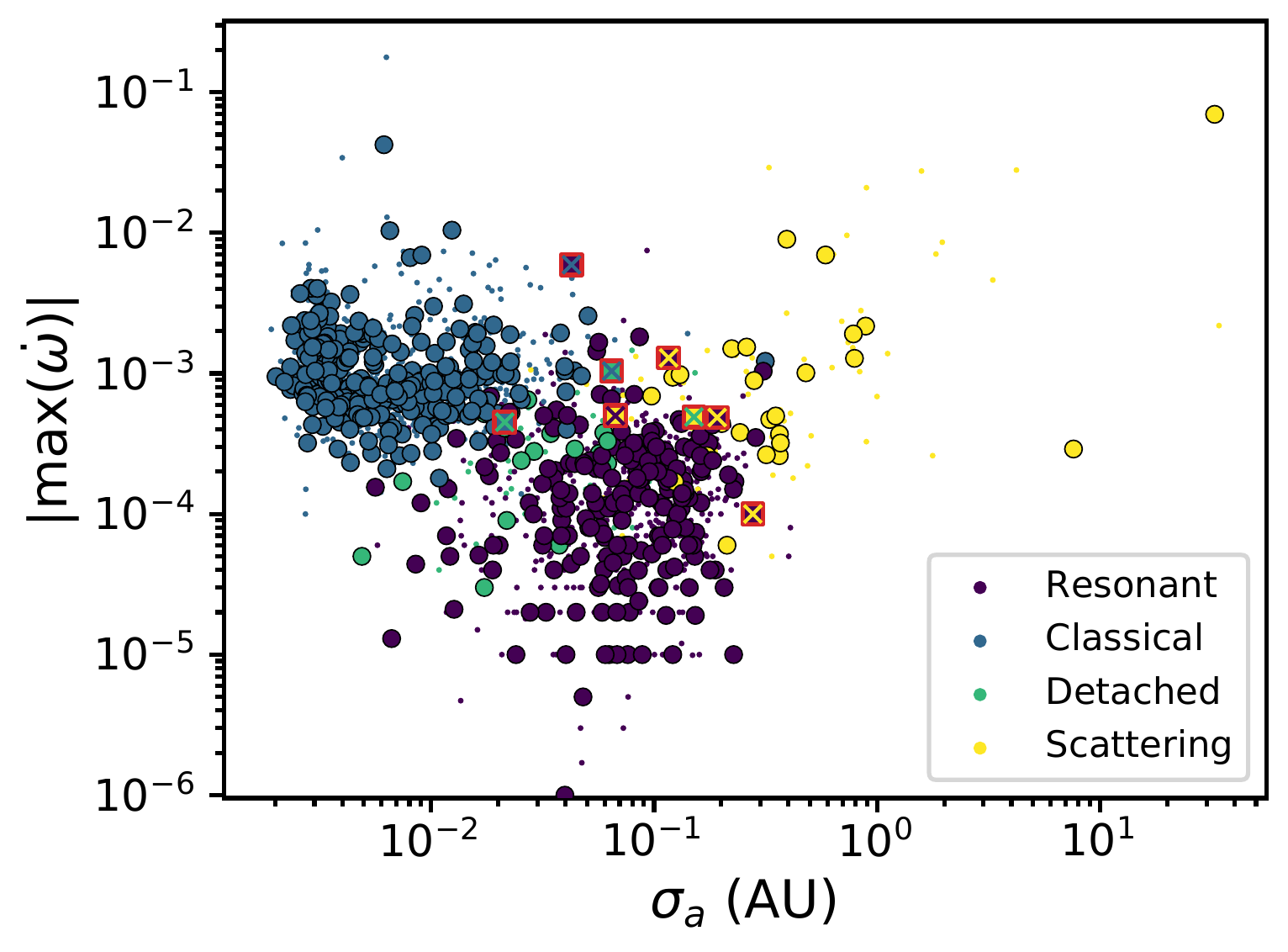} 
\caption{ The parameter space defined by the two most important features, $\sigma_a$ and $\max \dot{\omega}$. Small points show objects from the training set, outlined circles depict objects from the testing set, and squares outlined in red show the misclassified objects with the same convention as Figure~\ref{f:4class}. All of the misclassified objects lie along boundaries between classes. 
\label{f:important}}
\end{figure}

The performance of the best-fitting classifier is shown in Figure~\ref{f:4class}. 
Of the 542 objects in the testing set, only eight were misclassified. 
To dissect the physical intuition behind the classifier, we show the most important features that drive classification in Figure~\ref{f:features}. 
The two most important features for classification are the standard deviation in semi-major axis ($\sigma_\textrm{a}$) and the maximum time derivative of the argument of pericentre ($\max{\dot{\omega}}$). 
We show how objects occupy the parameter space for these two features in Figure~\ref{f:important}. 
Other parameters, including the spread of eccentricities and the changes in inclination, become important for further refinement of classes.  

As Figure~\ref{f:important} shows, the misclassified objects (identified by red boxes) typically lie along the boundary of multiple classes in feature space. Some of these objects are also near the boundaries between classes apparent in Figure~\ref{f:4class}; in several cases, the misclassified objects undergo late-simulation orbital evolution that changes the population the object belongs to in the 10\,Myr simulations. 
{Generally}, the classifier performs quite well on the majority of resonant objects despite the lack of semi-major axis normalization in the feature creation. 
In addition, the classifier is able to distinguish detached objects from classical KBOs despite the lack of a strong differentiation between the populations in terms of their orbital evolution (which is quite stable in both classes).

Figure~\ref{f:important} shows that, typically, classical objects have the smallest deviation in semi-major axis between the classes. The resonant objects have a much larger $\sigma_\textrm{a}$ because they undergo periodic changes in $a$ and $e$ as they librate in their resonances. Detached objects lie somewhere between classical and resonant. 
 Scattering objects have the largest $\sigma_\textrm{a}$ because of Neptune's influence on their orbital evolution. 
Classical objects tend to have a large, positive $\max{\dot{\omega}}$, while detached and resonant objects have a smaller value (or even negative value, for $\unsim20$ resonant KBOs).
{The longitude of perihelion, $\varpi$, is a sum of the argument of perihelion $\omega$ and the longitude of ascending node $\Omega$. For KBOs that are not strongly influenced by mean motion or secular resonances, $\varpi$ precesses and $\Omega$ regresses at rates determined by the net gravitational influence of the time-averaged orbits of the giant planets (see the discussion of secular theory in \citealt{Murray:1999}). These rates generally decrease with increasing semi-major axis with the exception of a sparsely populated region $a=40-42$~au in the classical belt where there are secular resonances \citep[see, e.g.,][]{Chiang:2008}.
Thus it makes sense that the relatively small range of semi-major axes for the classical KBOs translates into a relatively well-defined range of $\max{\dot{\omega}}$ in Figure~\ref{f:important}, and the more distant detached objects generally have lower values of $\max{\dot{\omega}}$.
For the resonant objects, the evolution of $\omega$ is influenced by resonant dynamics; the very small or negative values of $\max{\dot{\omega}}$ for some objects likely reflect so-called `Kozai' libration within mean motion resonances \citep[e.g.][]{Morbidelli:1993}, where $\dot{\Omega}$ and $\dot{\varpi}$ cancel, leading to libration of $\omega$. 
}
For scattering objects, $\max{\dot{\omega}}$ varies broadly, reflecting their wider range of semi-major axes and that their orbits undergo significant changes. 
Reassuringly, all of the misclassified objects lie upon a boundary between populations in Figure~\ref{f:important}, indicating that there is a physical ambiguity in the object's orbital evolution that the classifier is picking out. 
 
\begin{figure*}
\centering
\includegraphics[width=1.25\columnwidth]{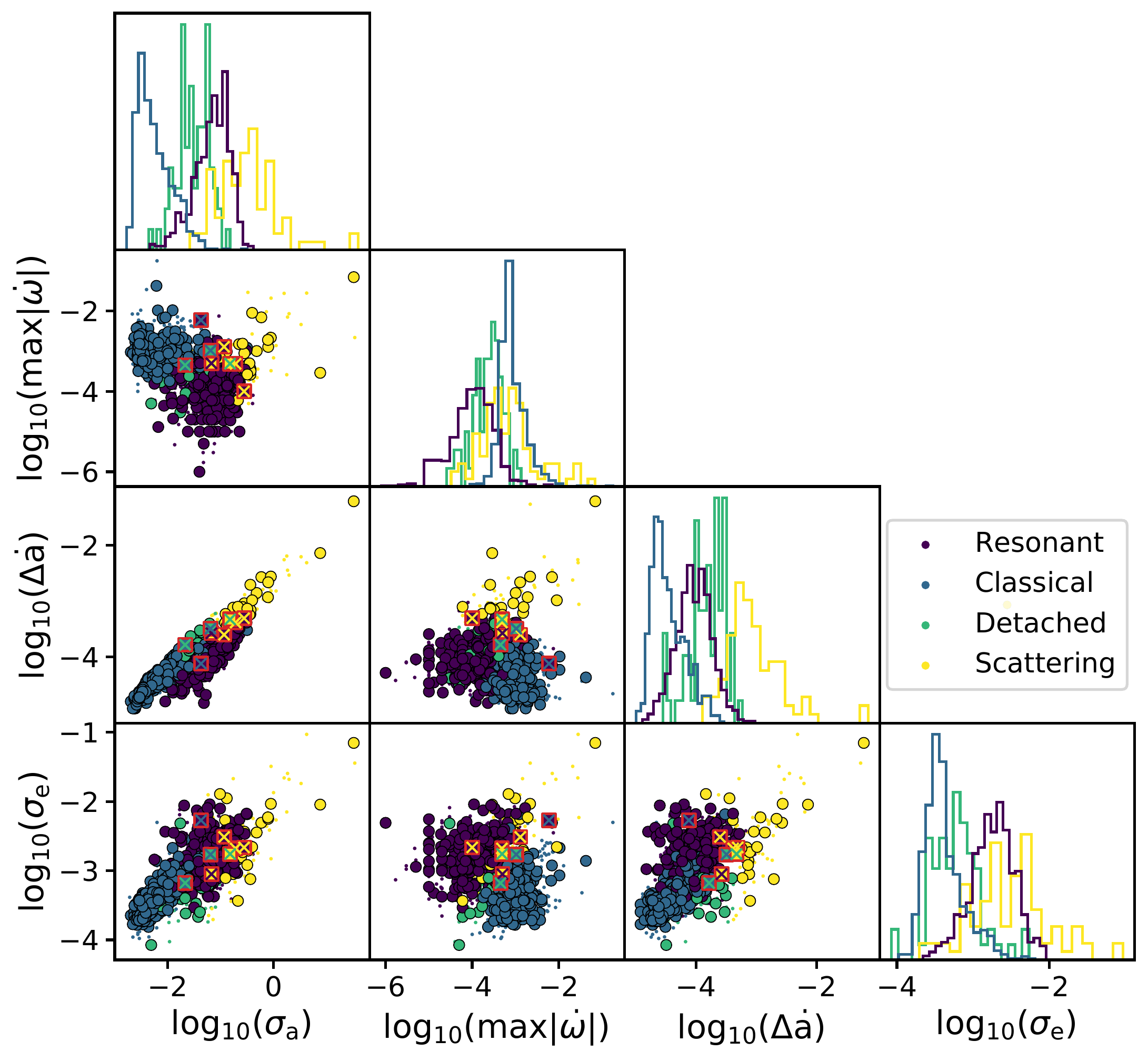} 
\caption{ Correlation plot for the four most important features in the classification algorithm. The plotting convention follows that of Figure~\ref{f:important}. Plots along the diagonal show normalized histograms of objects in the four classes.
\label{f:corner}}
\end{figure*}

We show distributions of and correlations between the top four features ($\sigma_\textrm{a}$, $\max{\dot{\omega}}$, $\Delta\dot{\textrm{a}}$, and $\sigma_\textrm{e}$) in Figure~\ref{f:corner}. Again, in most of these parameter spaces, the misclassified objects fall along a boundary between different classes. Similarly, the misclassified objects are never absolute extrema; rather, they tend to have middling values. None of the distributions for different classes are well-separated {in any individual features}, and the ordering of the peaks changes. This demonstrates that a two-parameter correlation is unable to uniquely separate the four classes, but a highly-multidimensional classification, such as is constructed by the classifier, is able to pick out nuances that lead to accurate classifications.

\subsection{Probability of Class Membership}\label{ss:probability}

\begin{figure}
\centering
\includegraphics[width=\columnwidth]{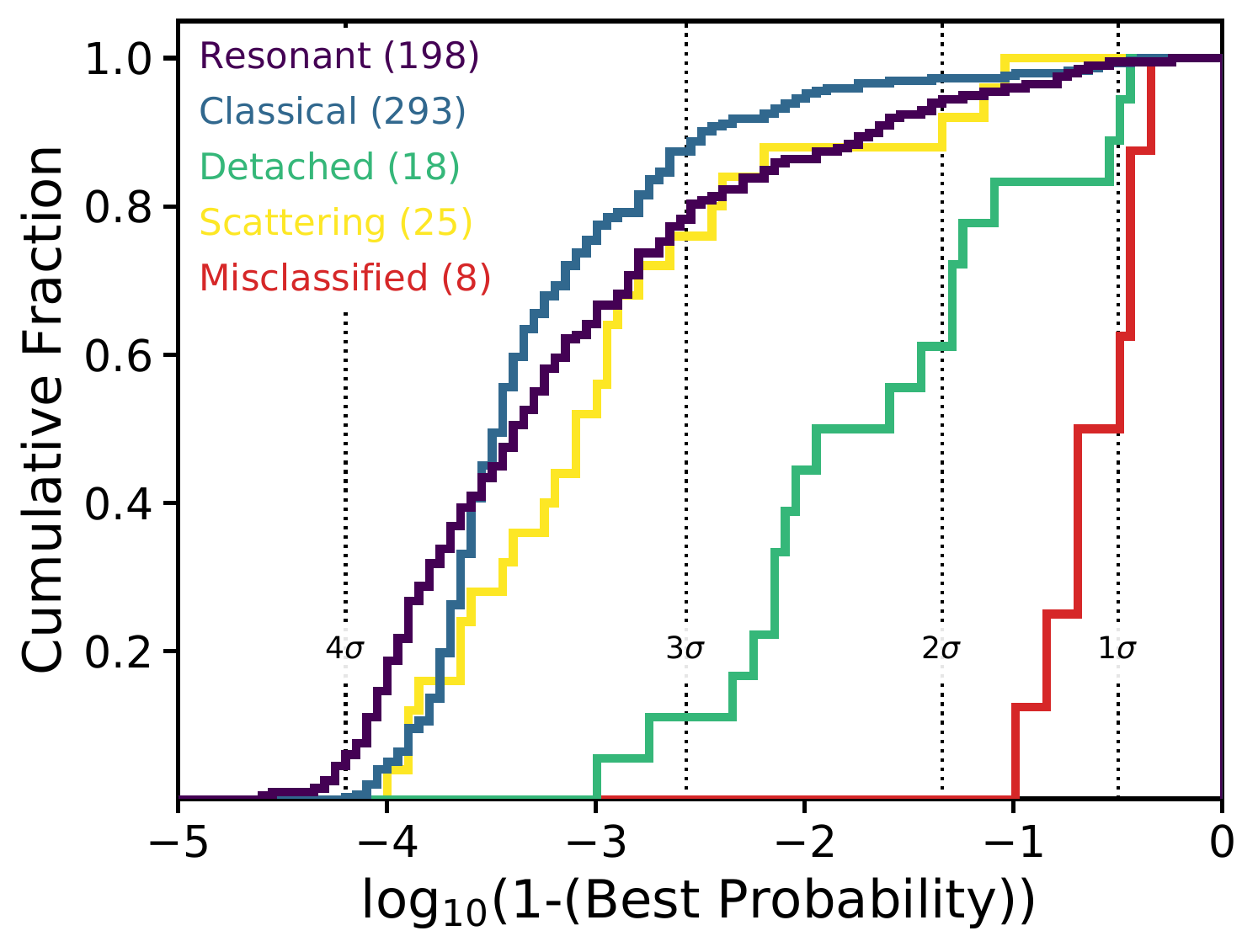}
\caption{Cumulative distribution of the probability of class membership for objects correctly identified in the four classes and for the misclassified objects. Vertical lines denote traditional confidence intervals. Over 80 per cent of objects boast a $>3\sigma$ probability of class membership, and all of the misclassified objects have a low probability of membership compared to others. This means that only a small fraction of objects could require human intervention for correct classification.
\label{f:prob}}
\end{figure}

For an automated method of object classification to be viable, a majority of objects should have a high probability of {correct} class membership, {and ideally all of the misclassified objects will have low class membership probabilities}. In the classification algorithm, each object is assigned a probability of membership for each of the classes such that the sum of all individual probabilities is one, and the assigned class is that with the highest probability.  We show the probabilities of class membership for the four populations of correctly classified objects and the misclassified set in Figure~\ref{f:prob}. Most objects, especially the common resonant and classical KBOs, have high probabilities of membership. Over 80 per cent of the testing set has a greater than $3\sigma$ probability of class membership: 79 per cent of resonant KBOs, 88 per cent of classical KBOs, 11 per cent of detached, and 75 per cent of scattering objects have very high probabilities of belonging to the correct population. 
{The detached objects have lower probabilities when compared to the other classes because there are relatively few detached objects with secure classifications, so they are under-represented in the training set. Many of the observed `detached' objects are either near the edges of mean motion resonances with Neptune or have orbit-fit uncertainties that encompass these resonances, thereby leading to lower class membership probabilities.} 
None of the misclassified objects has a probability greater than $2\sigma$. {By using even a conservative probability of membership cutoff of $3\sigma\approx99.7$ per cent,} we may be able to reduce the burden of human intervention substantially should a method like this be incorporated into a KBO classification pipeline. {We explore this further in Section~\ref{s:discussion}.}

\begin{figure}
\centering
\includegraphics[width=\columnwidth]{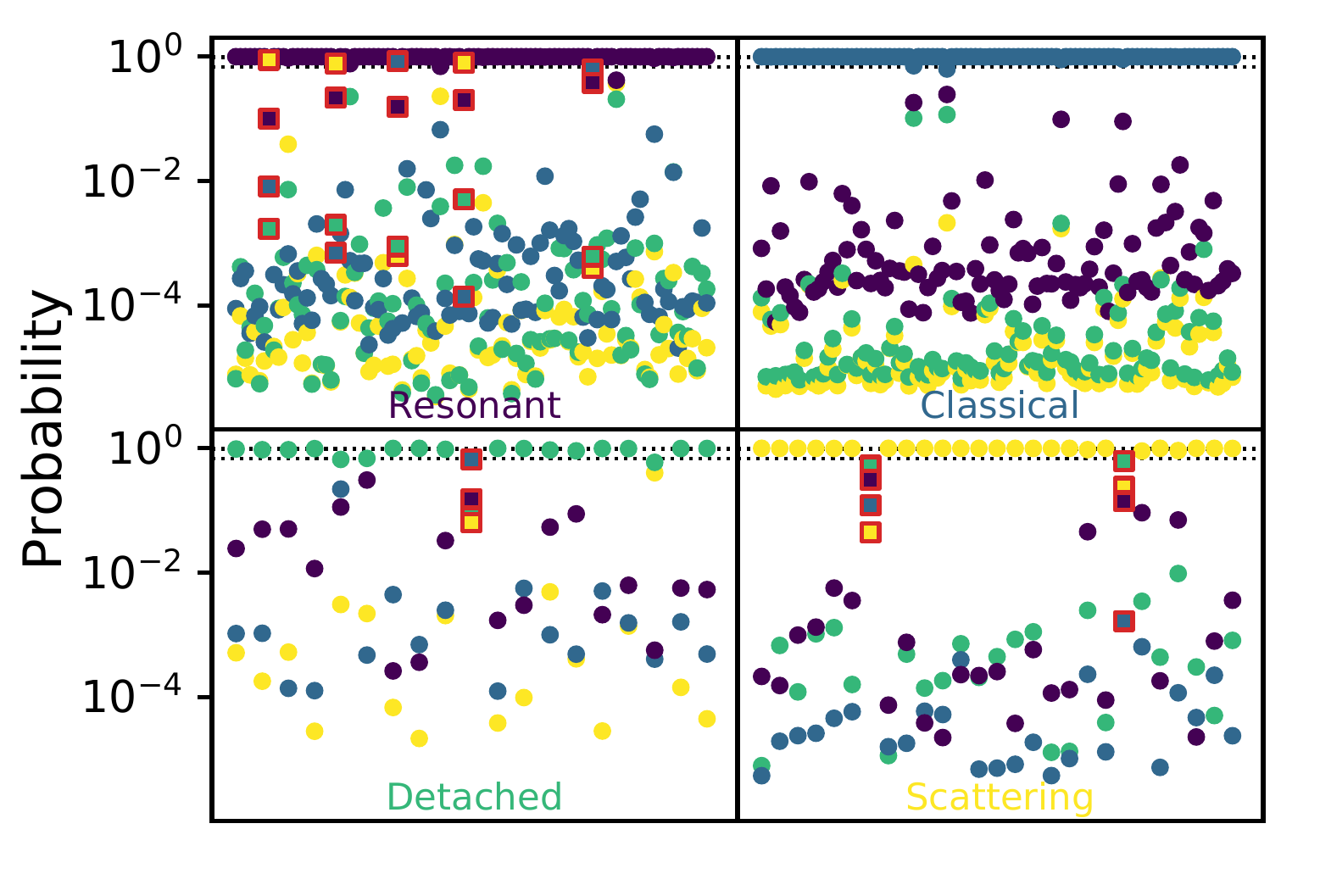}
\caption{{Probability of membership in all four classes for each object, with objects separated into quadrants based on their true class.} 
Each location along the horizontal axis {in a panel} shows a different object. An object's probability of membership for each of the four classes are shown as colour-coded dots { (which sum to one) } along that vertical line. Objects that were misclassified are identified by red squares. The number of correctly classified objects shown in the resonant and classical panels has been reduced to one hundred for visual clarity. From this, we can see that resonant objects and classical KBOs typically resemble each other the most due to the stratification of class probabilities, while detached and scattering objects don't have {an obvious trend for the ordering of other class probabilities}. 
\label{f:prob_scatter}}
\end{figure}

We can also examine the distribution of class probabilities for each object, which we show in Figure~\ref{f:prob_scatter}. 
The misclassified objects have a spread of class probabilities that is much smaller than other objects. Additionally, the class with the second highest preference is typically, but not always, the correct class, meaning that the classifier did pick up on some of the features associated with the correct classification. Most of the {test set objects} have well-stratified probabilities, indicating that one class is highly favored. {Figure~\ref{f:prob_scatter} shows that the} classical KBOs typically are very dominant in preference for their true class, {with resonant classification being consistently the next highest (though still low) probability.}
Classicals have a much lower probability of being detached or scattering.
Resonant, and scattering objects have little preference for the class with the second highest probability.
{There is a mild preference amongst the detached objects for a resonant classification being the next most probable. This is likely a reflection of the fact that the observed detached KBOs do tend to fall relatively close to resonances with Neptune.}

\subsection{Object Clone Classification}\label{ss:clones}

{The machine learning classifier presented above provides a more automated and computationally less expensive method to classify individual objects. As such, 
one particularly useful application of this type of algorithm is to make a more reliable classification pipeline with an expanded consideration of the orbit-fit uncertainties for each observed KBO.} A machine learning classifier could be used to classify `error clones' of an observed object that are initialized with perturbations in orbital elements drawn from the observational error space of the KBO.
Because we have used shorter numerical integrations (100\,kyr) in our classification scheme, we can run $\unsim100$ simulations of error clones for the same computational cost as the traditional 10\,Myr integration. Additionally, the inclusion of error clones in the classification provides better leverage on the probability of class membership: if the clones agree with the classification of the best fit, the object's classification can be better trusted.

\begin{figure}
\centering
\includegraphics[width=\columnwidth]{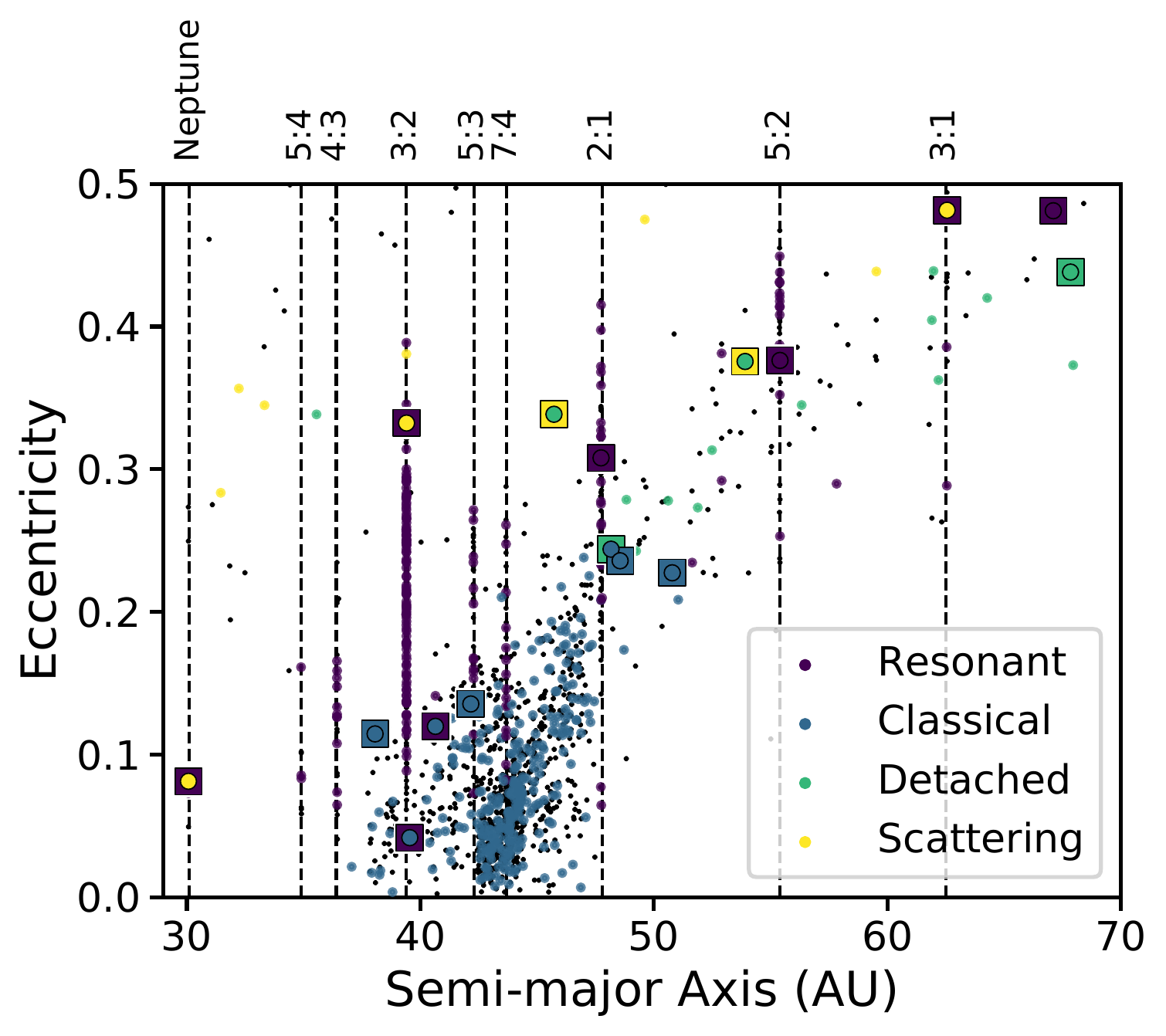}
\caption{Eccentricity vs. semi-major axis for clones of secure objects. Small black points in the background show the training data, while the small points show the clones where all three objects had the same classification and were correctly sorted into the true class. Squares show {KBOs in which at least one of the clones was not assigned the `true' class} ($\unsim3$ per cent of the testing data). The background colour of the square depicts the true class, while the coloured circle in the middle shows the class of the best-fitting orbit. 
\label{f:clones_ae}}
\end{figure}

We first investigate the classifications of the minimum and maximum error clones for each object {in the testing set} (drawn following the methodology of \cite{Gladman:2008} and described in Section~\ref{ss:classification}). {We call the collection of the independently classified minimum, best fit, and maximum clones for each observed KBO a `clone set' that we can analyse as an ensemble.} {Because our training and testing data draw only from the securely classified KBOs, these clones have the same true classifications as the best-fitting clones.} The error clones were classified using the fiducial GBC classification algorithm {trained on original training set of securely classified, best fit KBOs}. 
The classification of these objects is shown in Figure~\ref{f:clones_ae}. The majority of clones agreed with both the true class and the best fit class: 98 per cent of clones agreed with the best fit, and 97 per cent of clones agreed with the true value. There is no preferential population or area of parameter space in which the clone classification under-performs. 

\begin{figure}
\centering
\includegraphics[width=\columnwidth]{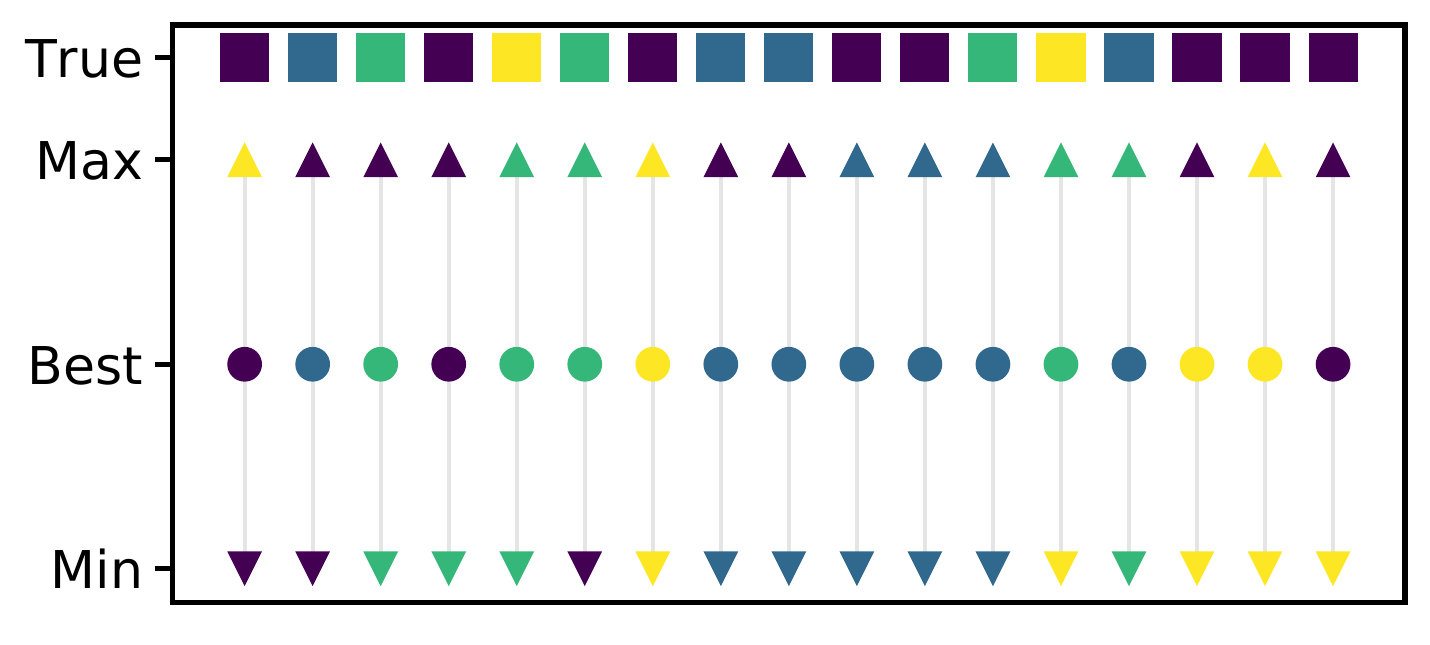}
\caption{ Comparison of the classification of the clones that did not agree with each other and/or the true value. Each column shows the true class (square), maximum error clone (up triangle), best fit orbit (circle), and minimum error clone (down triangle) for one object, with colours corresponding to the same classes used in previous plots, such as Figure~\ref{f:clones_ae}. We find objects where all three clones agree with each other but disagree with the true value {(e.g., second set from the right)}, and we find instances in which {at least two} clones disagree with each other {(e.g., left-most example)}. 
\label{f:clones_comp}}
\end{figure}

The 17 misclassified clone sets (out of 541 in the testing set) are explored further in Figure~\ref{f:clones_comp}. Six of the sets had all clones agree with each other {(meaning that the best fit and error clones were systematically misclassified)}, and 15 had at least one clone agree with the best fit. There is no preference for the type of clone disagreement seen: we find all populations mixed among the clones and true values.

\begin{figure}
\centering
\includegraphics[width=\columnwidth]{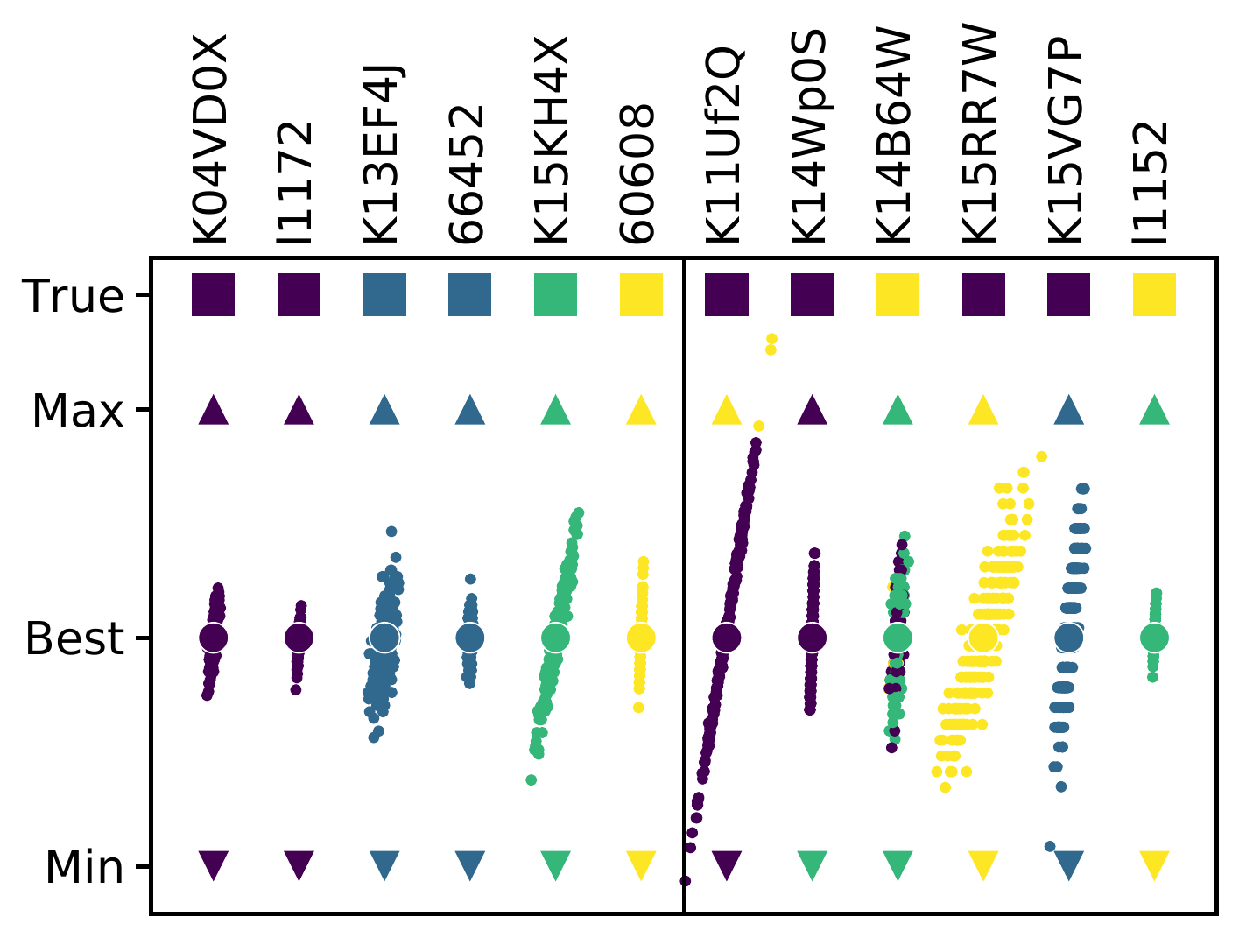}
\caption{ Clone classification for 250 clones drawn from the error space of the observations. The six objects on the left have secure classifications, where all clones agree with each other and the true classification, and the six objects on the left have clones that disagree with either each other or the true classification. The triangles, squares, and large circles are the same as in Figure~\ref{f:clones_comp}. The small points show the classification of individual clones. The vertical position of the point denotes the semi-major axis relative to the maximum and minimum error clones (triangles). The horizontal spread shows the eccentricity relative to the eccentricity of the best fit orbit (circle). Most of the objects on the right have a majority of clones agree with each other; K14B64W is the only object that has substantial disagreement between clones.
\label{f:errspace}}
\end{figure}

To better understand the behavior of clones in the full error space, we select six clone sets that agreed with both each other and the true classification and six clone sets that disagreed and ran simulations for an additional 250 clones {randomly sampled from the observational
uncertainties. To do this, we use the orbit uncertainties and corresponding covariance matrix calculated in the \citet{Bernstein:2000} orbit-fitting procedure to generate clones. 
This samples an uncertainty range somewhat different from the \cite{Gladman:2008} minimum and maximum clones because the covariance matrix does not account for potential systematic uncertainties in the astrometry reported for the objects.
The classifications of these clones} are shown in Figure~\ref{f:errspace}. 

{For illustrative purposes, we now discuss the individual behavior of the rightmost six objects from Figure~\ref{f:errspace}. } We find that, in most instances, only the extreme maximum and minimum error clones, or clones close to the extrema, differ from the best fit value. Four of six clone sets agree with the best fit value entirely. {The clone ensembles} also have very similar probabilities of class membership with one another (less than 0.2 dex for the examples shown here except for K14B64W\footnote{{We refer to individual objects by their packed designations; see \url{https://www.minorplanetcenter.net/iau/info/PackedDes.html}}}, which has a spread of nearly two orders of magnitude) and have similarly stratified probabilities in all classes (meaning that the probability ordering of one object's clones frequently does not change across the error space). 

K11Uf2Q only has three differing clones, which agree with the classification of the maximum error clone, { and the best fit agrees with the `true' classification. Similarly, K14Wp0S's clones agree with both the best fit value and the true classification, and only the minimum error clone disagrees with all other clones. Both of these cases show the power of using a large ensemble of clones: the machine learning method correctly identifies that the dynamical behavior of the vast majority of the ensemble is consistent, with only a few clones at the edges of the uncertainty range showing different {physical evolution.}
This provides a strong constraint on the true classification of the observed object. } 

{The error clone classifications of K15RR7W, K15VG7P, and l1152 all consistently disagree with the {`true'} classifications of the objects. K15VG7P is in a mixed-argument resonance, and the classifier incorrectly identifies it as a classical object. K15RR7W is a Neptune Trojan (in the 1:1 mean motion resonance with Neptune). The classifier did not perform well at identifying {low eccentricity} Trojans as resonant objects. 
We discuss the underlying reasons for these kinds of misclassifications in Section~\ref{ss:reasons}. KBO l1152 is a {weakly ($\unsim2$\,au)} scattering object on longer time-scales {($\unsim0.5$\,Myr)}, but has a very stable semi-major axis on the short time-scale integration {used by the classifier; therefore, the classifier correctly identified the population based on the information used but failed relative to the full 10\,Myr integration}. These three objects highlight the fact that these kinds of errors in the machine learning classifier will need to be addressed and/or characterized to be able to see the full benefits of error clone analysis. We discuss possible avenues of improvement in Section~\ref{ss:otherfeatures}.}

Only one object, K14B64W has substantial disagreement in clone classification {across the clone ensemble}. K14B64W's clones are classified as 36 per cent resonant, 57 per cent detached, and 8 per cent scattering. There is no obvious patterning of clone classification in semi-major axis, eccentricity, or inclination space; the identified class appears to be randomly distributed. {However, visual examination of the clones show variable dynamical evolution. This object intermittently librates in a high-order resonance. Thus, the ambiguity in the machine learning classification of clones reflects the diversity of dynamical evolution across the uncertainty range for this observed orbit.}

\section{Discussion}\label{s:discussion}

{Here we discuss in more detail the dominant reasons for errors in the machine learning classifier, both in the data set classified above and in two additional data sets to which we have applied the trained classifier. We suggest some future improvements that could be made to increase the accuracy of machine learning as applied to dynamical classification in the outer solar system.}

\subsection{Reasons for Misclassification}\label{ss:reasons}

\begin{figure}
\centering
\includegraphics[width=\columnwidth]{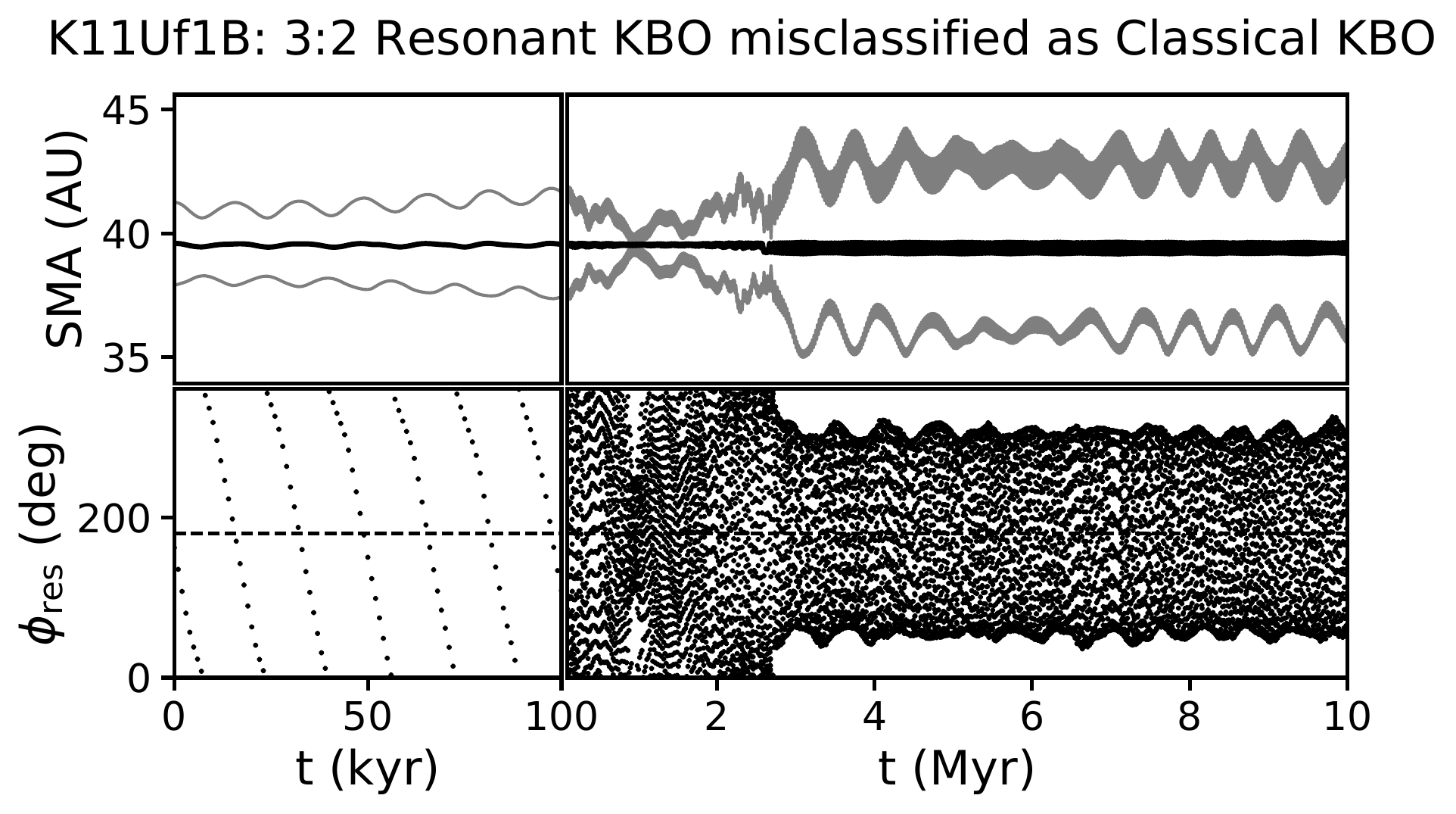}
\caption{ Semi-major axis and resonant angle as a function of time for a misclassified object. In the top panels, the black line shows the semi-major axis and the grey lines show the pericentre and apocentre distance. The bottom panels show the resonant angle (assuming libration in the 3:2 resonance) with the centre of libration ($180^\circ$) indicated by the horizontal dashed line. The left panels show the simulation used in the classifier ($10^5$ yr), while the right panel shows the simulation used to compute the true class ($10^7$ yr). This object shows one of the major reasons for misclassification: a dynamical event happens past the time of the simulation fed to the classifier (in this case, the object scatters into the 3:2 resonance at around 3\,Myr). The classifier correctly identified this object as a classical KBO given the short simulation it was given, but the long-term dynamical behavior clearly indicates that the object exists in the 3:2 resonance.
\label{f:whywrong}}
\end{figure}

{We examined in detail many of the cases of misclassification by the GBC classifier to get a sense of the dominant reasons for misclassification. 
Some of the misclassified objects are KBOs whose orbital evolution is just inherently ambiguous. These include cases where a judgement call was made to determine the `true' classification; examples of this include objects on the border between scattering and detached (i.e., semi-major axis variations very close to the empirical limit of 1.5~au determined by \citealt{Gladman:2008}). In other ambiguous cases, the true classification is resonant because an object's resonant angle librates for the majority of the 10\,Myr simulation, but that libration is intermittent. Some of these intermittently resonant objects are not librating at the very start of the integrations, so the machine learning classifier is not wrong, strictly speaking, when it classifies those as non-resonant based on short integrations; {Figure~\ref{f:whywrong} shows an example of this kind of `misclassification'.} 
{We see a few similar instances of misclassification of true scattering objects as detached where the object's short-term orbital evolution is quite stable, but longer integrations show it will scatter on 10\,Myr time-scales.
{In these cases, the use of the shorter integration time-scale results in a different classification than for longer integrations because the dynamical behavior changes significantly over time. 
There are other instances, however, when the short-time-scale behavior can predict the classification on longer time-scales, even if the short-time-scale behavior does not meet the \citet{Gladman:2008} definitions; an example of this is when the machine learning classifier correctly identifies a scattering object even though its semimajor axis does not undergo significant changes in the 100\,kyr integrations.}
While the classifications based on shorter integrations perform well overall compared to the 10\,Myr integrations in the \citet{Gladman:2008} scheme, this kind of behavior does highlight that classifications can be time-dependent.}

Other cases of misclassification are likely due to limitations of the training set. 
{For instance, the} algorithm is not always able to distinguish between classical and detached objects. This is likely partly because there are relatively few detached objects in the data set on which to train and partly because the boundary between classical and detached in the \citet{Gladman:2008} scheme is slightly arbitrary at smaller semi-major axes where these two populations share similar current orbital evolution.

In the misclassifications {described} above, the classifier tends to assign
{more equal probabilities to two or more classes, and the wrong `best' class is typically not strongly favored.}
The one instance we find in our data set where the classifier assigns a high probability to the wrong class membership is for resonant objects librating in mixed eccentricity and inclination type resonances. The vast majority of resonant KBOs discovered to date librate in eccentricity-type mean motion resonances, meaning that the resonant angle involves the KBO's longitude of perihelion and the libration causes coupled variations in the KBO's semi-major axis and eccentricity. However, there are a few {($\unsim10$)} resonant KBOs that have librating resonant angles that involve both the KBO's longitude of perihelion and the longitude of ascending node. These mixed-type resonances are generally weaker than the eccentricity-type resonances, so the variations in semi-major axis and eccentricity are less pronounced{; there are also variations in inclination not seen in eccentricity-type resonances}. The classifier, {which is essentially trained only on }eccentricity-type resonances, {therefore does not have the statistical power to recognize}
this alternative form of resonance. 
{The dominance of eccentricity-type resonances in the training set likely also contributes to the classifier's poor performance identifying low-eccentricity Neptune Trojans. The 1:1 resonant argument does not involve the longitude of perihelion, so it does not produce strongly coupled variations in $a$ and $e$ like most of the resonant population.}
We discuss how these insights could lead to improved machine learning classifiers in
Section~\ref{ss:otherfeatures}.}

\subsubsection{Classification of insecure objects}\label{sss:insecure}

{In the analysis presented above, we only show the classification of securely classified KBOs. We now investigate the performance of the fiducial classifier (trained on the secure objects) on the best-fitting orbits of the 500 KBOs from Table~\ref{T:cat} with insecure classifications according to the \citet{Gladman:2008} scheme.} {
The overall accuracy {for this data set} was unsurprisingly lower than for the secure objects at 75 per cent.  {Additionally, the probabilities of class membership were somewhat lower. Only about 70 per cent of correctly classified classical KBOs had more than a $3\sigma$ probability of class membership, and the other three classes contained $<40$ per cent correctly classified objects with high probability. }

{A reasonable fraction ($\unsim15$ per cent) of misclassified insecure KBOs possessed a high ($3\sigma$) probability of class membership. }
Of the incorrect classifications, we closely examined the 50 KBOs {(about 40 per cent of misclassified objects)} with (incorrect) classification probabilities at $>95$ per cent confidence. Of these, 6 are KBOs whose orbital evolution dances along the borders between classes{; for the integration provided to the classifier, the algorithm made a reasonable choice.} {Another 7 are mixed-argument} resonant objects that were incorrectly deemed classical due to the lack of representative objects in the {training} sample. There are 13 resonant KBOs that are only intermittently resonant and are not librating over the 100\,kyr time-scale integrations {and were thus misclassified}. Finally, 7 KBOs had `true' classifications that differ from the behavior of the best-fitting orbit{: the \cite{Gladman:2008} procedure assigns classifications based on the minimum and maximum clones in cases where those both agree even if they \emph{disagree} with the best-fitting orbit, meaning that the GBC classifier correctly identified the class of the best-fitting orbit. Thus, we have only} 17 KBOs that were unambiguously {misclassified}. 
Most of those 17 KBOs {have true classifications that place them in } high-order resonances ({which are} inherently more rare in the training data set) and/or have {libration} at very large amplitude within their resonances (meaning they are nearly non-resonant). Overall this is consistent with the reasons for misclassification found in the main data set. 
}

\begin{figure}
\centering
\includegraphics[width=\columnwidth]{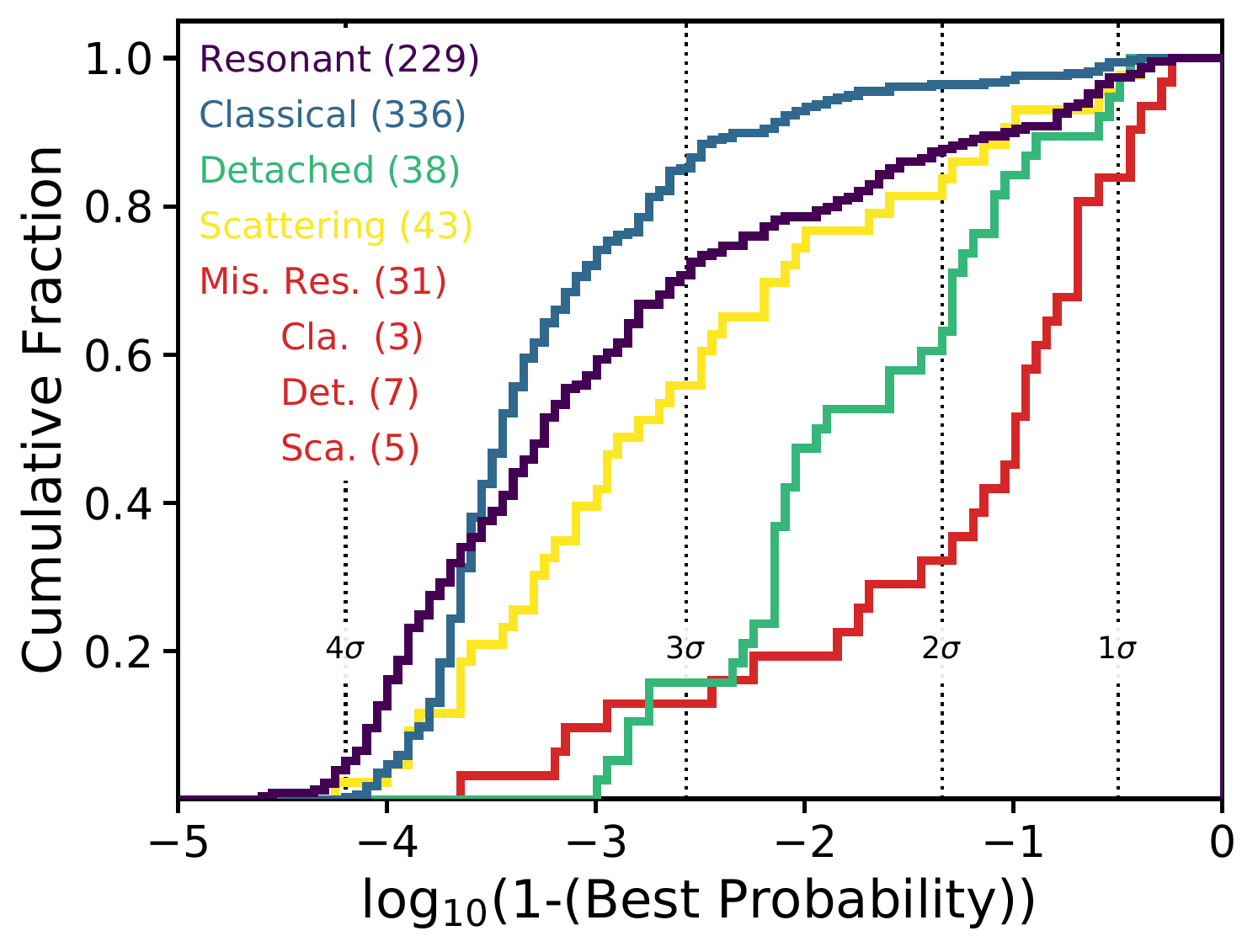}
\caption{{ Cumulative probability distributions for the combined secure and insecure sample (representing $30$ per cent of our overall KBO sample, with $\unsim80$ per cent secure and $\unsim20$ per cent insecure). 
The plot follows the conventions of Figure~\ref{f:prob}. Many of the misclassified objects with high probability are misclassified due the differences in time-scales between the 10\,Myr \protect\cite{Gladman:2008} classification scheme and the 100\,kyr method presented herein: the dynamics observed on shorter time-scales are consistent with a different class than the dynamics seen on longer time-scales.}
\label{f:insecure}}
\end{figure}

{
A typical observed set of KBOs will have a mix of secure and insecure classifications, so we can estimate the expected probability distribution for our classifier by mixing insecure objects into our test set.
Our overall set of KBOs has $\unsim20$ per cent insecure classifications (see Table~\ref{T:cat}), so adding a randomly chosen set of 135 insecurely classified KBOs to our testing set of 542 securely classified objects results in a typical mix of secure and insecure objects.
The expected combined probability distributions from the machine learning classifier for this set of KBOs is shown in Figure~\ref{f:insecure}.
The features we noted in Section~\ref{ss:probability}---a large fraction of high probabilities for correctly classified objects and a large fraction of low probabilities for misclassified objects---still hold true, which bodes well for the future applicability of this methodology. About 75 per cent of correctly classified objects have $>3\sigma$ probability of class membership, while more than $70$ per cent of misclassified objects have probabilities less than $2\sigma$. The small tail of high probability misclassifications results from objects suffering from classification time-scale ambiguity as described above.  }

\subsection{Classification of DECam objects}\label{ss:decam}

The recent publication of observations of 131 new KBOs discovered using the Cerro Tololo-DECam \citep{Wasserman2020} provides an independent set of objects on which to test the trained {machine learning} algorithm. 
Using the astrometry provided {by the Minor Planet Center}, we followed the \citet{Gladman:2008} procedure described in Section~\ref{ss:classification} to determine the `true' classifications of these new KBOs. {From this procedure, we find 54 classical, 52 resonant, 18 detached, and 7 scattering KBOs in this data set.}

\begin{figure}
\centering
\includegraphics[width=\columnwidth]{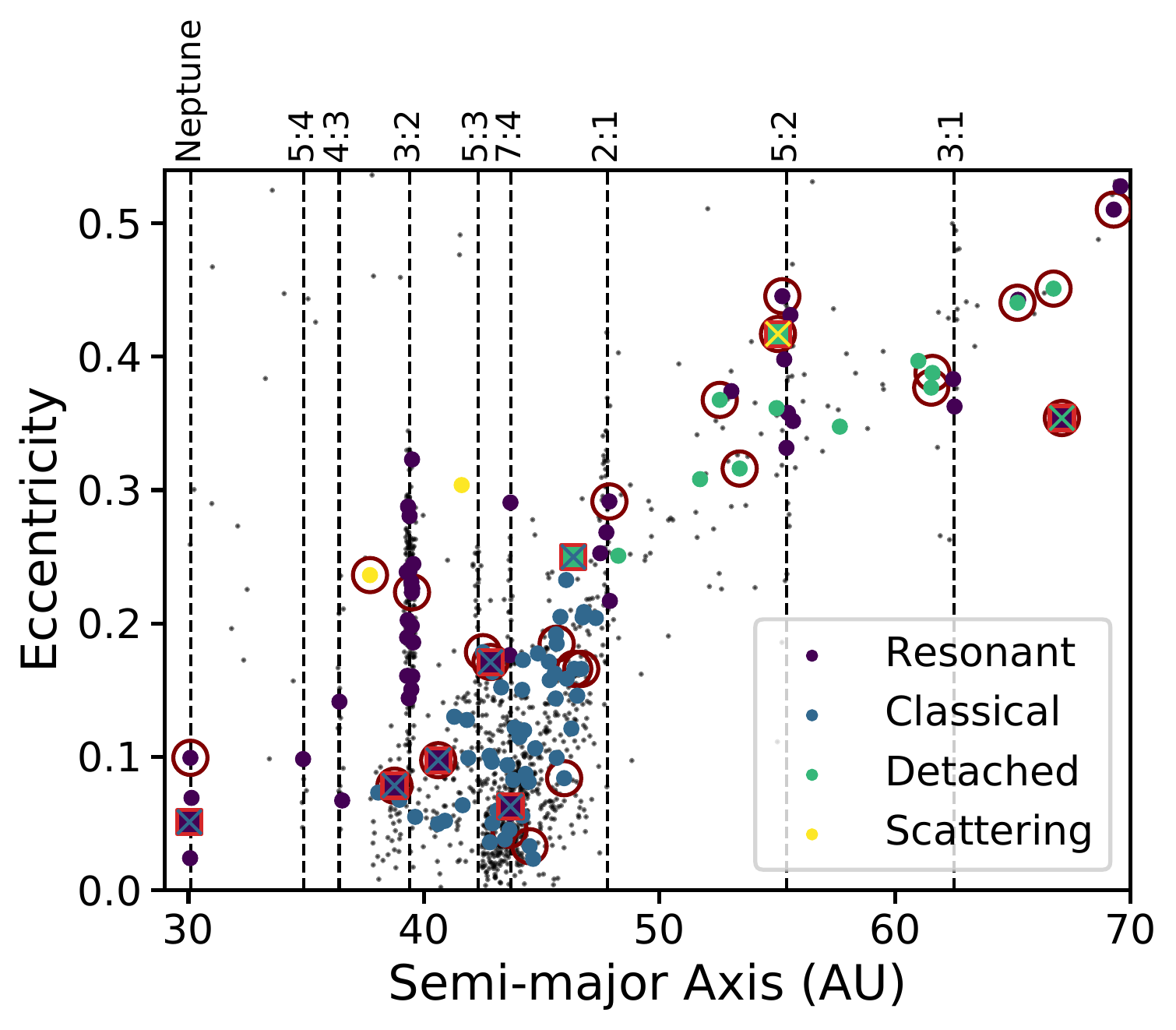}
\caption{Eccentricity vs. semi-major axis for the DECam objects following the colouring conventions of Figure~\ref{f:4class}. Maroon circles surround objects that have an insecure classification by the traditional method of classification {(8 of the insecure objects are outside the plot limits)}. Of the 131 DECam objects classified here, 10 disagreed with the labelled class (92.4 per cent accuracy) despite 33 of those labelled classifications being insecure.
\label{f:decam_ae}}
\end{figure}

We then classify all of the DECam data using the GBC classifier developed above (which uses the same training set of objects as the classifier in Section~\ref{s:results}). We do not remove objects with insecure classifications from the data set, as we aim to characterize the performance of the classifier on `unknown' data {with a fairly typical mix of secure and insecurely classifiable objects}, as would happen if this methodology was folded into a blind pipeline analysis. Without any modifications to the algorithm, the machine learning classifications agreed with the true classifications 92.4 per cent of the time, as shown in Figure~\ref{f:decam_ae}. We have 10 objects that do not agree with the true classification, and seven of those are objects that have an insecure true classification. The misclassifications in this data set occupy a similar parameter space to misclassifications in our fiducial testing set: the classifier finds ambiguity in a few low eccentricity resonant objects and an object that is close to the classical--detached boundary. 

\begin{figure}
\centering
\includegraphics[width=\columnwidth]{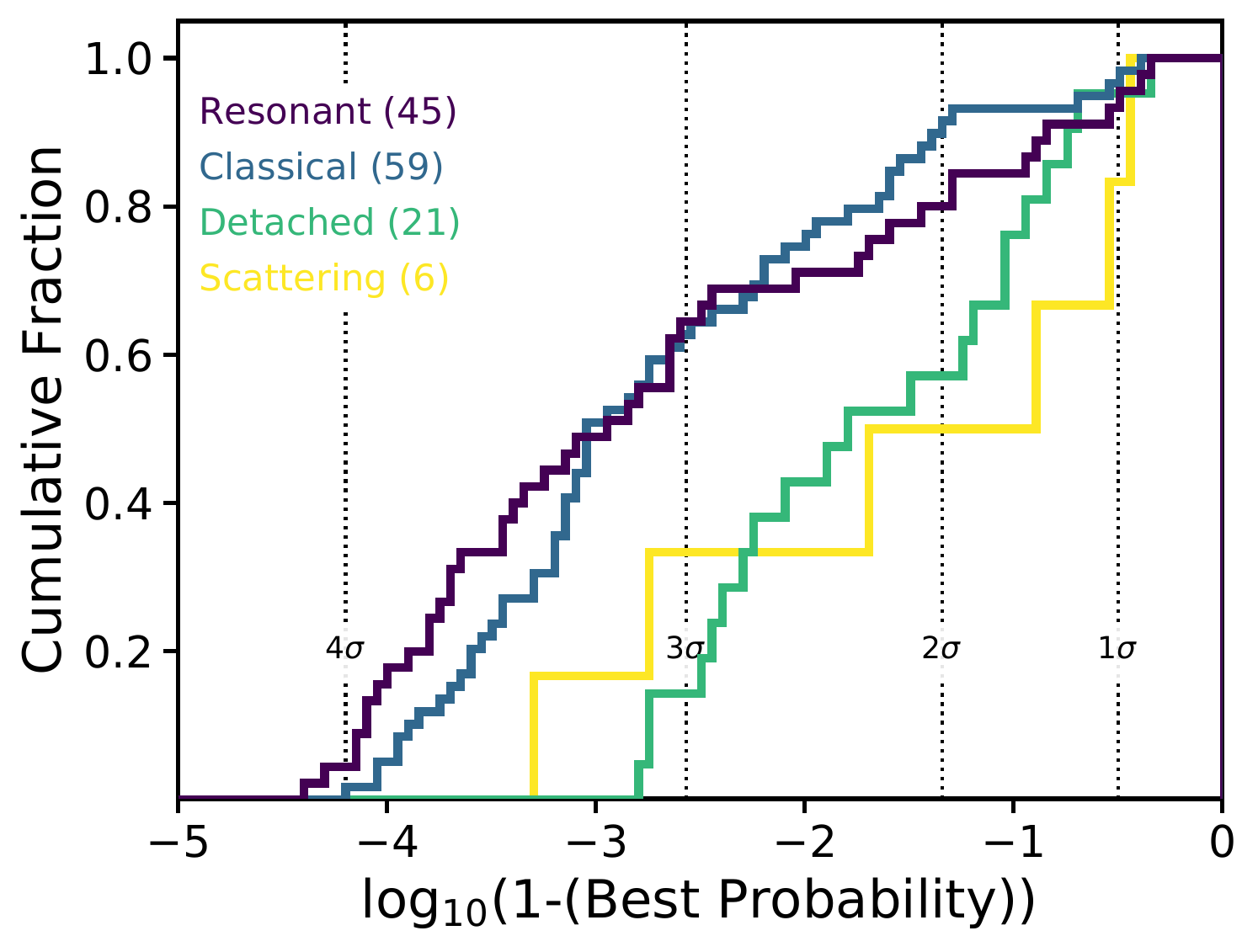}
\caption{ The cumulative probabilities of class membership for the DECam objects based on the class assigned by {the GBC machine learning classifier}. Over 50 per cent of objects have a $>3\sigma$ probability of class membership despite the relatively large error of the observed orbits that act as inputs to this method.
\label{f:decam_prob}}
\end{figure}

We show the best-fitting class probability in Figure~\ref{f:decam_prob}. In this figure, we do not differentiate between the correct classifications and misclassified objects in an effort to simulate the distribution that might be expected from a blind classification of unknown data. Similarly to the fiducial data set {and the joint secure/insecure analysis in Section~\ref{sss:insecure}}, a majority of classical (63 per cent) and resonant (62 per cent) objects have best-fitting probabilities of $>3\sigma$. Only one of the ten misclassified objects (K13RC4O, a true resonant classified as a classical) is a `false positive' with a best-fitting probability of $3.00\sigma$; {this object is in a mixed-argument resonance, and the true classification is insecure}. 

Because this data set is very manageable in size, we visually examined the dynamical evolution of each clone of the 131 new KBOs to determine exactly how well the {machine learning} algorithm did and to identify features of the objects classified incorrectly or classified at low probability.
We find that every false classification made at $>90$ per cent confidence was the result of the true classification being a mixed-argument resonance. 
The majority of both the {misclassifications} and true classifications made at lower probabilities are {either} resonant objects with low eccentricities or very small semi-major axis libration or are objects whose `correct' classifications are a judgment call because they are on the border between different dynamical classes. 
{The algorithm generally gives lower probabilities to the detached classifications. An examination of the `true' detached KBOs in this data set reveals that they are mostly insecure classifications; 15 of the 18 detached KBOs are close to the boundaries of resonances with Neptune, and an additional 2 nearly have enough semi-major axis mobility on 10\,Myr time-scales to be classified as scattering rather than detached. Thus, the lower probabilities for class membership are appropriate. }

\subsection{Improvements to the classifier}\label{ss:otherfeatures}

\subsubsection{Alternate Features}
In this work, we have used only the features that are simplest to directly extract from the numerical simulations {in the machine learning classifiers}. 
However, one could imagine that more dynamically-motivated features, such as pericentre distance, could improve classifications, especially of objects like the Neptune Trojans (objects in the 1:1 resonance) or scattering objects that cross Neptune's orbit. To test if new features improves the classification, we compute the pericentre distance for all objects at each simulation step and extract the same set of features described in Section~\ref{sss:features} (mean, standard deviation, time derivative, etc.) to add to the feature ensemble used to train the classifier. We also add {some additional features for the orbital angle evolution}: we include the total slope of $\omega$ and $\Omega$, which are intended to be less sensitive to {short time-scale} variations than $\dot{\omega}$. Using the same set of objects in the training and testing sets, albeit with the new features, we achieve the same accuracy as found for the fiducial data. The best-fitting probability distributions are nearly identical; the cumulative fractions vary by less than 0.07 across all bins for all classes. The ordering and relative influence of feature importances are the only change resulting from the addition of new features. The second most important feature in the fiducial data, $\max{\dot{\omega}}$, is replaced by the slope of $\omega$, and the initial pericentre distance becomes the third most important feature. However, because pericentre is a higher-dimensional combination of features that already exist in the data, little \emph{new} information is added to the classifier, leading to no better accuracy in object classification.

\subsubsection{Synthetic Kuiper Belt Catalogs}

The classifier developed herein has the most difficulty classifying less common objects, such as objects in mixed-argument resonances. {The dominant} factor in the classifier's struggle with these types of objects is low number statistics. Because only a few examples are known across the entire Kuiper Belt, there aren't sufficient numbers in the training set for the classifier to extract a strong set of {features} that characterizes the population well. To alleviate this issue, one could create a more balanced data set (meaning that there are more equal numbers in different populations) using simulated objects. The benefits of a synthetic catalog include an extremely well-characterized input sample and the ability to include large numbers of objects in every population class. {The larger number of objects in a synthetic data set would also allow separation of some classes into sub-components. This would be particularly useful for identifying different kinds of resonant behavior. If the mixed-argument resonant objects, which have different characteristic $a$, $e$, and $i$ evolution compared to the eccentricity-type resonant objects, were included in the classifier as an independent class, the classifier would likely achieve a better and more accurate classification for this population. This approach might also help with classifying intermittently resonant objects; feeding the classifier a labelled set of objects that are not librating during the 100\,kyr integrations but that will librate for the majority of a 10\,Myr integration might help it identify features associated with being on the boundary of a resonance.}
{Just as with an observed data set, a} synthetic data set runs the risk that any type of object not included in the synthetic data set (such as a type of resonance not yet seen) may not be properly classified; {but conversely, in a synthetic data set, one could include theoretically hypothesized or predicted} populations of objects not yet observed. A synthetic catalog might also be more useful for {sub-classifying resonant objects into their specific resonances rather than into one bulk class of resonant objects.} This idea will be explored in a future paper. 

\section{Conclusions}\label{s:conclusions}

In this paper, we successfully demonstrate that very short ($\unsim100$\,kyr) numerical integrations can be used to classify observed KBOs into their dynamical populations (resonant, classical, detached, and scattering) with high accuracy ($>90$ per cent). We use 1805 KBOs that have secure classifications as the fiducial data set and take the traditional 10\,Myr classification as the `true' classification; the data consist of 642 resonant, 998 classical, 74 detached, and 91 scattering KBOs. We find the following.

\begin{enumerate}
  \item A gradient boosting tree classifier, which is a multi-class machine learning classifier that uses several initializations and layers of sorting trees for classification, achieves a 98 per cent accuracy on the fiducial {testing set of 542 securely classified KBOs}. We find that the most important data features for classification are the standard deviation of semi-major axis ($\sigma_a$) and the maximum time derivative of the argument of pericentre ($\max\dot{\omega}$). 
  
  \item We find that $\unsim80$ per cent of our {fiducial} testing set has a $>3\sigma$ probability of belonging to the correct class. Misclassified objects typically have a much lower probability of class membership ($<2\sigma$), making them easy to differentiate. {Almost all of the objects assigned higher probabilities of incorrect classifications (i.e., potential false positives) are objects that are in mixed-argument mean motion resonances with Neptune; this is an uncommon type of resonance inadequately sampled in our training set.}
  
  \item {Misclassified objects tend to lie along the boundaries between populations in feature space, indicating that the classifier is picking up on ambiguity between dynamical populations. Many of these objects have orbital evolution that makes their true classifications difficult to determine.}
 
  \item {Objects drawn from the observed error space of known KBOs typically have the same classification and probability of class membership as the best fit.  Because the machine learning algorithm operates on a shorter integration time-scale than the fiducial 10\,Myr classifications, a promising avenue for future classification pipelines is to use those computational savings to better explore the uncertainty range of the observed KBOs' orbit fits. We demonstrate how a large ensemble of clones can be used to better verify population membership for an object and to provide insight into the intrinsic physical variability of an orbit's observed error distribution.  }

  \item We test our algorithm on a completely new data set of 131 KBOs recently released from a DECam survey \citep{Wasserman2020} and find a 92 per cent agreement with the `true' classifications despite only 75 per cent of the objects having a secure true classification. This successful performance on an unknown data set {with a mix of secure and and insecure `true' classifications that is typical for surveys} indicates that a machine learning classifier will be viable for any new observed KBOs.
\end{enumerate}

We have shown that a simple machine learning classifier will be a viable and valuable tool for classifying KBOs in the LSST era. We are able to substantially reduce both the computational and human resources needed to label observed KBOs into their dynamical populations, which will be critical as the number of objects grows by an order of magnitude in the next several years. {Methods like this have also recently been proven successful for asteroid family members \citep{Carruba2020}. There are a number of improvements to the work presented here that would enable this methodology to be more reliable for expected large survey observations. Most importantly, a large synthetic training set should be created; such a training set would allow for more accurate classification of rarely-observed dynamical types, prepare for classification of as-yet unobserved object types, and perhaps enable for a more detailed division of dynamical classes (e.g., classifying into individual resonances).}  {Being able to accurately and efficiently classify Kuiper Belt observations will enable science by allowing for detailed comparisons of planet formation and evolution models with the Kuiper Belt today.}

\section*{Acknowledgements}
We thank our reviewer Jean-Marc Petit for his thoughtful analysis of our work. We thank Leon Palafox for input on an early iteration of this project, and we thank Kaitlin Kratter for helpful comments on this manuscript. RAS acknowledges support from the National Science Foundation under Grant No. DGE-1143953. KV acknowledges support from NSF grant AST-1824869 and NASA grants NNX14AG93G, NNX15AH59G, and 80NSSC19K0785.

\section*{Data Availability}
The data features for all best fit orbits in the sample and the classifier developed in this article can be found at \url{https://github.com/rsmullen/KBO_Classifier}.  Other data in this work will be shared on reasonable request to the corresponding author.



\bibliographystyle{mnras}
\bibliography{references} 

\begin{thebibliography}{}
\makeatletter
\relax
\def\mn@urlcharsother{\let\do\@makeother \do\$\do\&\do\#\do\^\do\_\do\%\do\~}
\def\mn@doi{\begingroup\mn@urlcharsother \@ifnextchar [ {\mn@doi@}
  {\mn@doi@[]}}
\def\mn@doi@[#1]#2{\def\@tempa{#1}\ifx\@tempa\@empty \href
  {http://dx.doi.org/#2} {doi:#2}\else \href {http://dx.doi.org/#2} {#1}\fi
  \endgroup}
\def\mn@eprint#1#2{\mn@eprint@#1:#2::\@nil}
\def\mn@eprint@arXiv#1{\href {http://arxiv.org/abs/#1} {{\tt arXiv:#1}}}
\def\mn@eprint@dblp#1{\href {http://dblp.uni-trier.de/rec/bibtex/#1.xml}
  {dblp:#1}}
\def\mn@eprint@#1:#2:#3:#4\@nil{\def\@tempa {#1}\def\@tempb {#2}\def\@tempc
  {#3}\ifx \@tempc \@empty \let \@tempc \@tempb \let \@tempb \@tempa \fi \ifx
  \@tempb \@empty \def\@tempb {arXiv}\fi \@ifundefined
  {mn@eprint@\@tempb}{\@tempb:\@tempc}{\expandafter \expandafter \csname
  mn@eprint@\@tempb\endcsname \expandafter{\@tempc}}}

\bibitem[\protect\citeauthoryear{Adorf \& Meurs}{Adorf \&
  Meurs}{1988}]{Adorf1988}
Adorf H.~M.,  Meurs E. J.~A.,  1988, in Seitter W.~C.,  Duerbeck H.~W.,   Tacke
  M.,  eds, Large-{{Scale Structures}} in the {{Universe Observational}} and
  {{Analytical Methods}}. Lecture {{Notes}} in {{Physics}}.
{Springer}, {Berlin, Heidelberg}, pp 315--322,
  \mn@doi{10.1007/3-540-50135-5_86}

\bibitem[\protect\citeauthoryear{{Bannister} et~al.,}{{Bannister}
  et~al.}{2018}]{Bannister:2018}
{Bannister} M.~T.,  et~al., 2018, \mn@doi [\apjs] {10.3847/1538-4365/aab77a},
  \href {https://ui.adsabs.harvard.edu/abs/2018ApJS..236...18B} {236, 18}

\bibitem[\protect\citeauthoryear{{Bernstein} \& {Khushalani}}{{Bernstein} \&
  {Khushalani}}{2000}]{Bernstein:2000}
{Bernstein} G.,  {Khushalani} B.,  2000, \mn@doi [\aj] {10.1086/316868}, \href
  {https://ui.adsabs.harvard.edu/abs/2000AJ....120.3323B} {120, 3323}

\bibitem[\protect\citeauthoryear{{Brasser}, {Duncan}, {Levison}, {Schwamb}  \&
  {Brown}}{{Brasser} et~al.}{2012}]{Brasser:2012}
{Brasser} R.,  {Duncan} M.~J.,  {Levison} H.~F.,  {Schwamb} M.~E.,   {Brown}
  M.~E.,  2012, \mn@doi [\icarus] {10.1016/j.icarus.2011.10.012}, \href
  {https://ui.adsabs.harvard.edu/abs/2012Icar..217....1B} {217, 1}

\bibitem[\protect\citeauthoryear{Carruba, Aljbaae, Domingos, Lucchini  \&
  Furlaneto}{Carruba et~al.}{2020}]{Carruba2020}
Carruba V.,  Aljbaae S.,  Domingos R.~C.,  Lucchini A.,   Furlaneto P.,  2020,
  \mn@doi [Monthly Notices of the Royal Astronomical Society]
  {10.1093/mnras/staa1463}

\bibitem[\protect\citeauthoryear{Chen et~al.,}{Chen et~al.}{2019}]{Chen2019}
Chen H. H.-H.,  et~al., 2019, \mn@doi [ApJ] {10.3847/1538-4357/ab1a40}, 877, 93

\bibitem[\protect\citeauthoryear{{Chiang} \& {Choi}}{{Chiang} \&
  {Choi}}{2008}]{Chiang:2008}
{Chiang} E.,  {Choi} H.,  2008, \mn@doi [\aj] {10.1088/0004-6256/136/1/350},
  \href {https://ui.adsabs.harvard.edu/abs/2008AJ....136..350C} {136, 350}

\bibitem[\protect\citeauthoryear{Choudhary, Lindner, Holliday, Miller, Sinha
  \& Ditto}{Choudhary et~al.}{2019}]{Choudhary2019}
Choudhary A.,  Lindner J.~F.,  Holliday E.~G.,  Miller S.~T.,  Sinha S.,
  Ditto W.~L.,  2019, arXiv:1912.01958 [physics]

\bibitem[\protect\citeauthoryear{{Dones}, {Brasser}, {Kaib}  \&
  {Rickman}}{{Dones} et~al.}{2015}]{Dones:2015}
{Dones} L.,  {Brasser} R.,  {Kaib} N.,   {Rickman} H.,  2015, \mn@doi [\ssr]
  {10.1007/s11214-015-0223-2}, \href
  {https://ui.adsabs.harvard.edu/abs/2015SSRv..197..191D} {197, 191}

\bibitem[\protect\citeauthoryear{{Elliot} et~al.,}{{Elliot}
  et~al.}{2005}]{Elliot:2005}
{Elliot} J.~L.,  et~al., 2005, \mn@doi [\aj] {10.1086/427395}, \href
  {https://ui.adsabs.harvard.edu/abs/2005AJ....129.1117E} {129, 1117}

\bibitem[\protect\citeauthoryear{{Gladman} \& {Chan}}{{Gladman} \&
  {Chan}}{2006}]{Gladman:2006}
{Gladman} B.,  {Chan} C.,  2006, \mn@doi [\apjl] {10.1086/505214}, \href
  {https://ui.adsabs.harvard.edu/abs/2006ApJ...643L.135G} {643, L135}

\bibitem[\protect\citeauthoryear{{Gladman}, {Holman}, {Grav}, {Kavelaars},
  {Nicholson}, {Aksnes}  \& {Petit}}{{Gladman} et~al.}{2002}]{Gladman:2002}
{Gladman} B.,  {Holman} M.,  {Grav} T.,  {Kavelaars} J.,  {Nicholson} P.,
  {Aksnes} K.,   {Petit} J.~M.,  2002, \mn@doi [\icarus]
  {10.1006/icar.2002.6860}, \href
  {https://ui.adsabs.harvard.edu/abs/2002Icar..157..269G} {157, 269}

\bibitem[\protect\citeauthoryear{{Gladman}, {Marsden}  \&
  {Vanlaerhoven}}{{Gladman} et~al.}{2008}]{Gladman:2008}
{Gladman} B.,  {Marsden} B.~G.,   {Vanlaerhoven} C.,  2008, {Nomenclature in
  the Outer Solar System}.
pp 43--57

\bibitem[\protect\citeauthoryear{{Ivezi{\'c}} \& et al.}{{Ivezi{\'c}} \&
  et~al.}{2019}]{Ivezic:2019}
{Ivezi{\'c}} {\v{Z}}.,  et al. 2019, \mn@doi [\apj] {10.3847/1538-4357/ab042c},
  \href {https://ui.adsabs.harvard.edu/abs/2019ApJ...873..111I} {873, 111}

\bibitem[\protect\citeauthoryear{{Kaib}, {Ro{\v{s}}kar}  \& {Quinn}}{{Kaib}
  et~al.}{2011}]{Kaib:2011}
{Kaib} N.~A.,  {Ro{\v{s}}kar} R.,   {Quinn} T.,  2011, \mn@doi [\icarus]
  {10.1016/j.icarus.2011.07.037}, \href
  {https://ui.adsabs.harvard.edu/abs/2011Icar..215..491K} {215, 491}

\bibitem[\protect\citeauthoryear{Lam \& Kipping}{Lam \&
  Kipping}{2018}]{Lam2018}
Lam C.,  Kipping D.,  2018, \mn@doi [MNRAS] {10.1093/mnras/sty022}, 476, 5692

\bibitem[\protect\citeauthoryear{{Lawler} et~al.,}{{Lawler}
  et~al.}{2019}]{Lawler:2019}
{Lawler} S.~M.,  et~al., 2019, \mn@doi [\aj] {10.3847/1538-3881/ab1c4c}, \href
  {https://ui.adsabs.harvard.edu/abs/2019AJ....157..253L} {157, 253}

\bibitem[\protect\citeauthoryear{{Levison} \& {Duncan}}{{Levison} \&
  {Duncan}}{1994}]{Levison:1994}
{Levison} H.~F.,  {Duncan} M.~J.,  1994, \mn@doi [\icarus]
  {10.1006/icar.1994.1039}, \href
  {https://ui.adsabs.harvard.edu/abs/1994Icar..108...18L} {108, 18}

\bibitem[\protect\citeauthoryear{{Lykawka} \& {Mukai}}{{Lykawka} \&
  {Mukai}}{2007}]{Lykawka:2007}
{Lykawka} P.~S.,  {Mukai} T.,  2007, \mn@doi [\icarus]
  {10.1016/j.icarus.2007.06.007}, \href
  {https://ui.adsabs.harvard.edu/abs/2007Icar..192..238L} {192, 238}

\bibitem[\protect\citeauthoryear{{Malhotra}}{{Malhotra}}{2019}]{Malhotra:2019}
{Malhotra} R.,  2019, \mn@doi [Geoscience Letters] {10.1186/s40562-019-0142-2},
  \href {https://ui.adsabs.harvard.edu/abs/2019GSL.....6...12M} {6, 12}

\bibitem[\protect\citeauthoryear{{McKinnon} et~al.,}{{McKinnon}
  et~al.}{2020}]{McKinnon:2020}
{McKinnon} W.~B.,  et~al., 2020, \mn@doi [Science] {10.1126/science.aay6620},
  \href {https://ui.adsabs.harvard.edu/abs/2020Sci...367.6620M} {367, aay6620}

\bibitem[\protect\citeauthoryear{McLeod, Libeskind, Lahav  \& Hoffman}{McLeod
  et~al.}{2017}]{McLeod2017}
McLeod M.,  Libeskind N.,  Lahav O.,   Hoffman Y.,  2017, \mn@doi [JCAP]
  {10.1088/1475-7516/2017/12/034}, 2017, 034

\bibitem[\protect\citeauthoryear{{Morbidelli} \& {Moons}}{{Morbidelli} \&
  {Moons}}{1993}]{Morbidelli:1993}
{Morbidelli} A.,  {Moons} M.,  1993, \mn@doi [\icarus]
  {10.1006/icar.1993.1052}, \href
  {https://ui.adsabs.harvard.edu/abs/1993Icar..102..316M} {102, 316}

\bibitem[\protect\citeauthoryear{{Morbidelli}, {Levison}  \&
  {Gomes}}{{Morbidelli} et~al.}{2008}]{Morbidelli:2008}
{Morbidelli} A.,  {Levison} H.~F.,   {Gomes} R.,  2008, {The Dynamical
  Structure of the Kuiper Belt and Its Primordial Origin}.
pp 275--292

\bibitem[\protect\citeauthoryear{Murray \& Dermott}{Murray \&
  Dermott}{1999a}]{Murray1999}
Murray C.~D.,  Dermott S.~F.,  1999a, Solar {{System Dynamics}}.
{Cambridge University Press}, {Cambridge, UK}

\bibitem[\protect\citeauthoryear{{Murray} \& {Dermott}}{{Murray} \&
  {Dermott}}{1999b}]{Murray:1999}
{Murray} C.~D.,  {Dermott} S.~F.,  1999b, {Solar system dynamics}.
{Cambridge University Press}

\bibitem[\protect\citeauthoryear{{Nesvorn{\'y}}}{{Nesvorn{\'y}}}{2015}]{Nesvorny:2015}
{Nesvorn{\'y}} D.,  2015, \mn@doi [\aj] {10.1088/0004-6256/150/3/73}, \href
  {https://ui.adsabs.harvard.edu/abs/2015AJ....150...73N} {150, 73}

\bibitem[\protect\citeauthoryear{{Nesvorn{\'y}}, {Li}, {Youdin}, {Simon}  \&
  {Grundy}}{{Nesvorn{\'y}} et~al.}{2019a}]{Nesvorny:2019b}
{Nesvorn{\'y}} D.,  {Li} R.,  {Youdin} A.~N.,  {Simon} J.~B.,   {Grundy} W.~M.,
   2019a, \mn@doi [Nature Astronomy] {10.1038/s41550-019-0806-z}, \href
  {https://ui.adsabs.harvard.edu/abs/2019NatAs...3..808N} {3, 808}

\bibitem[\protect\citeauthoryear{{Nesvorn{\'y}} et~al.,}{{Nesvorn{\'y}}
  et~al.}{2019b}]{Nesvorny:2019}
{Nesvorn{\'y}} D.,  et~al., 2019b, \mn@doi [\aj] {10.3847/1538-3881/ab3651},
  \href {https://ui.adsabs.harvard.edu/abs/2019AJ....158..132N} {158, 132}

\bibitem[\protect\citeauthoryear{Nord, Connolly, Kinney, Kubica, Narayan, Peek,
  Schafer  \& Tollerud}{Nord et~al.}{2019}]{Nord2019}
Nord B.,  Connolly A.~J.,  Kinney J.,  Kubica J.,  Narayan G.,  Peek J. E.~G.,
  Schafer C.,   Tollerud E.~J.,  2019, BAAS, 51, 224

\bibitem[\protect\citeauthoryear{Pedregosa et~al.,}{Pedregosa
  et~al.}{2011}]{Pedregosa2011a}
Pedregosa F.,  et~al., 2011, Journal of Machine Learning Research, 12, 2825

\bibitem[\protect\citeauthoryear{{Pike} et~al.,}{{Pike}
  et~al.}{2017}]{Pike:2017}
{Pike} R.~E.,  et~al., 2017, \mn@doi [\aj] {10.3847/1538-3881/aa83b1}, \href
  {https://ui.adsabs.harvard.edu/abs/2017AJ....154..101P} {154, 101}

\bibitem[\protect\citeauthoryear{Schwamb et~al.,}{Schwamb
  et~al.}{2018}]{Schwamb2018}
Schwamb M.~E.,  et~al., 2018, arXiv:1802.01783 [astro-ph]

\bibitem[\protect\citeauthoryear{{Schwamb} et~al.,}{{Schwamb}
  et~al.}{2019}]{Schwamb:2019}
{Schwamb} M.~E.,  et~al., 2019, \mn@doi [Research Notes of the American
  Astronomical Society] {10.3847/2515-5172/ab0e10}, \href
  {https://ui.adsabs.harvard.edu/abs/2019RNAAS...3...51S} {3, 51}

\bibitem[\protect\citeauthoryear{{Storrie-Lombardi}, Lahav, Sodr{\'e}  \&
  {Storrie-Lombardi}}{{Storrie-Lombardi} et~al.}{1992}]{Storrie-Lombardi1992}
{Storrie-Lombardi} M.~C.,  Lahav O.,  Sodr{\'e} L.,   {Storrie-Lombardi} L.~J.,
   1992, \mn@doi [MNRAS] {10.1093/mnras/259.1.8P}, 259, 8P

\bibitem[\protect\citeauthoryear{Tamayo et~al.,}{Tamayo
  et~al.}{2016}]{Tamayo2016}
Tamayo D.,  et~al., 2016, \mn@doi [ApJ] {10.3847/2041-8205/832/2/L22}, 832, L22

\bibitem[\protect\citeauthoryear{{Volk} \& {Malhotra}}{{Volk} \&
  {Malhotra}}{2017}]{Volk:2017}
{Volk} K.,  {Malhotra} R.,  2017, \mn@doi [\aj] {10.3847/1538-3881/aa79ff},
  \href {https://ui.adsabs.harvard.edu/abs/2017AJ....154...62V} {154, 62}

\bibitem[\protect\citeauthoryear{Wasserman et~al.,}{Wasserman
  et~al.}{2020}]{Wasserman2020}
Wasserman H.~L.,  et~al., 2020, Minor Planet Electronic Circulars, 2020-C88

\bibitem[\protect\citeauthoryear{{Yu}, {Murray-Clay}  \& {Volk}}{{Yu}
  et~al.}{2018}]{Yu:2018}
{Yu} T. Y.~M.,  {Murray-Clay} R.,   {Volk} K.,  2018, \mn@doi [\aj]
  {10.3847/1538-3881/aac6cd}, \href
  {https://ui.adsabs.harvard.edu/abs/2018AJ....156...33Y} {156, 33}

\makeatother
\end{thebibliography}






\bsp	
\label{lastpage}
\end{document}